\documentclass[10pt,twocolumn,showpacs,amsmath,amssymb, aps, prb]{revtex4-1}

\usepackage{graphicx}
\usepackage{dcolumn}
\usepackage{bm}
\usepackage[cp1250]{inputenc}
\usepackage[usenames,dvipsnames]{color}

\renewcommand{\today}{Submitted version of November 14, 2016} 

\begin{document}
\title{Optical phonons for Peierls chains with long-range Coulomb interactions}

\author{M\'at\'e Tim\'ar$^{1,2}$}
\author{Gergely Barcza$^1$}
\author{Florian Gebhard$^3$}
\email{florian.gebhard@physik.uni-marburg.de}
\author{\"Ors Legeza$^1$}
\email{legeza.ors@wigner.mta.hu}
\affiliation{$^1$Strongly Correlated Systems Lend\"ulet Research Group, 
Institute for Solid State Physics and Optics, MTA Wigner Research Centre for
Physics, P.O.\  Box 49, H-1525 Budapest, Hungary}
\affiliation{$^2$Department of Physics of Complex Systems, E\"otv\"os
University, H-1117 Budapest, Hungary}
\affiliation{$^3$Fachbereich Physik, Philipps-Universit\"at Marburg,
D-35032 Marburg, Germany}

\date{\today}

\begin{abstract}%
We consider a chain of atoms that are bound together by a harmonic force.
Spin-1/2 electrons that move between neighboring chain sites 
(H\"uckel model) induce a lattice dimerization
at half band filling (Peierls effect). We supplement the
H\"uckel model with a local Hubbard interaction and a long-range Ohno potential,
and calculate the average bond-length, dimerization, 
and optical phonon frequencies for finite straight and zig-zag chains
using the density-matrix renormalization group (DMRG) method.
We check our numerical approach against analytic results 
for the H\"uckel model.
The Hubbard interaction mildly affects the average bond length but
substantially enhances the dimerization and increases the optical phonon frequencies
whereas, for moderate Coulomb parameters, the long-range Ohno interaction 
plays no role. 
\end{abstract}



\maketitle

\section{Introduction}
\label{sec:Introduction}

The calculation of lattice vibrations in ordinary metals and band insulators
is one of the basic tasks in theoretical solid-state 
physics.~\cite{BornHuang,Bilz,Devreese,RevModPhys.73.515,Solyom}
Phonon dispersions can be measured with inelastic neutron 
scattering,~\cite{Dorner,Lovesey,Furrer}
and Raman and infrared spectroscopy permit the detection of
vibrations with finite energy and vanishingly 
small momenta (`optical phonons') in crystals 
and in molecules.~\cite{Claus,Schrader,Hackl}

For correlated electron systems, however,
the calculation of phonon frequencies is still at its beginning. 
For one-dimensional systems for example,
theoretical investigations focus on the ground-state phase diagram 
of the Holstein-Hubbard model
where the competition between the electron-phonon coupling and 
the electron-electron interaction leads to a rich ground-state phase
diagram.~\cite{PhysRevB.83.195105,Advances2010,%
PhysRevLett.109.116407,PhysRevB.91.085114,PhysRevB.91.235150}

In the adiabatic limit where the phonons can be treated classically,
one-dimensional electronic systems at half band-filling dimerize 
(Peierls effect),~\cite{Solyom2}
as is observed in $\pi$-conjugated polymers such as 
trans-polyacetylene.~\cite{Baeriswyl,Barford}
The phonon spectrum of such a Peierls insulator cannot be described by
short-range forces (`harmonic springs') acting between atoms at short distances
because the optical phonon branch shows a non-trivial momentum dependence
with a strong reduction at small momenta (Kohn anomaly);~\cite{Solyom2} 
a Peierls chain of non-interacting electrons provides 
a well-known example for the Kohn anomaly.~\cite{supplemental}
However, it is not well understood how the electron-electron interaction 
influences the (optical) phonon frequencies.

In this work, 
we study electrons that move between neighboring sites on a half-filled 
chain (H\"uckel model)~\cite{Salem}
so that the system describes a Peierls insulator.
We add a local Hubbard interaction and a long-range Ohno potential that 
approximates the electrons' Coulomb interaction
(H\"uckel-Hubbard and H\"uckel-Hubbard-Ohno models).
Reliable numerical investigations of ground-state and excited-state properties 
for long chains have become possible only recently
using the Density-Matrix Renormalization Group 
(DMRG) method;~\cite{White-1992,White-1993}
for developments of the method in the last decade,
see Refs.~[\onlinecite{Schollwock-2005,Schollwock-2011,
Szalay-2015}] and references therein.
We find that the Coulomb interaction suppresses the Kohn anomaly and 
increases the frequency of the optical phonons. 

Our work is structured as follows.
In Sect.~\ref{sec:model}, we set up our H\"uckel-Hubbard-Ohno Hamiltonian 
for the itinerant electrons that move over a straight or zig-zag
backbone of harmonically bound
atoms. Moreover, we define the backbone distortions that correspond to optical phonons.

In Sect.~\ref{sec:Hueckelmodel}, we analyze the H\"uckel model
for non-interacting electrons analytically for periodic boundary conditions
and numerically for open boundary conditions using the DMRG method.
For Peierls insulators with a sizable gap, the average bond length,
the dimerization, the single-particle gap, and the optical phonon frequencies
for systems with up to $L_C=110$ sites can safely be extrapolated 
to the thermodynamic limit.

In Sect.~\ref{sec:HHOmodel},
we show that the Hubbard interaction is primarily responsible for the
increase of the dimerization and of the phonon frequencies in comparison
with the results for the bare H\"uckel model.
A moderately large Ohno interaction leads to very small corrections 
only. Since the Hubbard interaction substantially increases the single-particle
gap, finite-size effects are well under control.
For our parameter set, the calculated optical phonon frequencies on a zig-zag chain 
are in the range of measured values in 
trans-polyacetylene.~\cite{Kuzmany,Takeuchi1,Takeuchi2,Ehrenfreund}

Summary and conclusions, Sect.~\ref{sec:summconc}, close our 
presentation. Technical details are deferred to two appendices and
the supplemental material.

\section{Model}
\label{sec:model}

Our model study mimics the properties of trans-polyacetylene.
We focus on the carbon-carbon stretch and bend modes so that we
can work with a small set of parameters. 
The calculation of optical phonon frequencies 
in trans-polyacetylene requires a more sophisticated
description of the structure, namely, the motion of the hydrogen atoms (C-H
bond stretching and bending) must be included.

\subsection{Structure}

\begin{figure}[hb]
\includegraphics[height=7.3cm,angle=90]{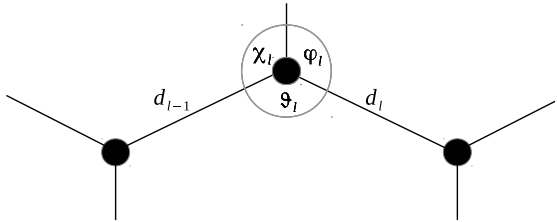}
\caption{Coordinates and angles in a planar and unflexed dimerized zig-zag 
chain.\label{fig:structure}}
\end{figure}

The carbon atoms in trans-polyacetylene are arranged in a zig-zag chain.
For a perfect sp$^2$ hybridization of the carbon $2s$-$2p$ orbitals, 
the atoms arrange in a zig-zag chain as ground-state conformation 
with $\vartheta_l^{(0)}=\chi_l^{(0)}=\varphi_l^{(0)}=\Theta_0=2\pi/3=120^{\circ}$,
see Fig.~\ref{fig:structure}.
For illustrative purposes and for
comparison with earlier work, we shall also address a straight chain
with $\vartheta_l^{(0)}=\pi=180^{\circ}, \chi_l^{(0)}=\varphi_l^{(0)}=\pi/2=90^{\circ}$.

For our analytic calculations we consider a chain 
with $L_C=2L$ atoms that is supposed to be planar and unflexed.
The atoms occupy the positions ($l=1,2,\ldots 2L$)
\begin{equation}
\vec{r}_l =
\left( \begin{array}{c}
x_l\\
y_l
\end{array}
\right) \; .
\end{equation}
We denote the ground-state coordinates by capital letters,
$\vec{r}_l^{\, (0)}\equiv (X_l,Y_l)^{\rm T}$.
We orient the chain to the right of the origin,
$X_1=Y_1=0$ and $X_3>0,Y_3=0$ in the ground state.
For our numerical investigations, we add two atoms, one at the beginning and one
at the end of the chain, whose positions are kept fixed during the geometry
optimization. 
In the following we formulate our equations for $L_C=2L$.

\subsection{Contributions to the ground-state energy}

In the ground state, the bond lengths are alternating between long `single' bonds
and short `double' bonds.
The Peierls distortion is  due to the itinerant electrons
that interact via the Hubbard-Ohno interaction.
For the zig-zag chain, a clock spring models
the repulsive interaction of the $\sigma$-bonds.

\subsubsection{Electronic Hamiltonian}

The system is half filled, i.e., the number of electrons 
equals the number of sites,
$N_{\uparrow}+N_{\downarrow}=L_C=2L$; it is paramagnetic, 
$N_{\uparrow}=N_{\downarrow}=L_C/2=L$.
The electrons move between neighboring sites (H\"uckel model)~\cite{Salem}
\begin{equation}
\hat{T}= -\sum_{\sigma}\sum_{l=1}^{2L-1}t_l
\left( 
\hat{c}_{l+1,\sigma}^{\dagger}\hat{c}_{l,\sigma}^{\vphantom{\dagger}} 
+\hat{c}_{l,\sigma}^{\dagger}\hat{c}_{l+1,\sigma}^{\vphantom{\dagger}} 
\right) \; ,
\label{eq:kinetic-operator}
\end{equation}
where $\hat{c}_{l,\sigma}^{\dagger}$ ($\hat{c}_{l,\sigma}^{\vphantom{\dagger}}$)
creates (annihilates) an electron with spin $\sigma=\{\uparrow,\downarrow\}$
on carbon atom~$l$. 
The parameters for the electron transfer between nearest neighbors
are given by the Peierls expression ($l=1,2,\ldots,2L-1$)
\begin{equation}
t_l= t(d_l)=t_0 \exp\left( -(d_l-r_0)\alpha/t_0\right) \;,
\label{eq:defti}
\end{equation}
where $t_0=2.5\, {\rm eV}$ is the electron transfer parameter at 
distance $r_0=1.4\, \hbox{\AA}$, and
$\alpha=4.0\, {\rm eV}/\hbox{\AA}$ parameterizes
the Peierls coupling.~\cite{PCCPlegeza}
Moreover, the nearest-neighbor distances $d_l$ as a function of the coordinates
$\left\{x_l\right\},\left\{y_l\right\}$ are given by
\begin{equation}
d_l=|\vec{d}_l|=\sqrt{(x_{l+1}-x_l)^2+(y_{l+1}-y_l)^2} \; .
\label{eq:dldef}
\end{equation}
More generally, we denote the distance between the atoms~$i$ and $j$ by
\begin{equation}
d_{ij}=|\vec{r}_i-\vec{r}_j| = \sqrt{(x_i-x_j)^2+(y_i-y_j)^2} \; .
\end{equation}

The Coulomb interaction between the electrons 
is given by the Hubbard-Ohno interaction,
\begin{eqnarray}
\hat{H}_{\rm int}&=&
U\sum_{l=1}^{2L}\left(\hat{n}_{l,\uparrow}-1/2\right)\left(\hat{n}_{l,\downarrow}-1/2\right)
\nonumber \\
&& + \frac{1}{2\epsilon_d}\sum_{l\neq m=1}^{2L}V(d_{lm})
\left(\hat{n}_l-1\right)\left(\hat{n}_m-1\right) \; ,
\label{eq:coulomb-operator}
\end{eqnarray}
where $\hat{n}_{l,\sigma}=\hat{c}_{l,\sigma}^{\dagger}
\hat{c}_{l,\sigma}^{\vphantom{\dagger}}$ counts the number of $\sigma$-electrons
on carbon atom~$l$, and $\hat{n}_l=\hat{n}_{l,\uparrow}+\hat{n}_{l,\downarrow}$.
We parameterize the distance-dependence of the density-density interaction
by the Ohno expression~\cite{Baeriswyl,Barford}
\begin{equation}
V(x)= \frac{V}{\sqrt{1+\beta(x/\hbox{\AA})^2}} \; ,\, 
\beta= \left(\frac{V}{14.397\, {\rm eV}}\right)^2 \; .
\label{eq:Ohnoexpression}
\end{equation}
The Ohno form guarantees that, at large distances, 
the electrons interact via their unscreened
Coulomb interaction, $e^2=14.397\, {\rm eV}\hbox{\AA}$.
In this study we use the Coulomb and screening parameters 
$U=6\, {\rm eV}$, $V=3\, {\rm eV}$, $\epsilon_d=2.3$, as in our
investigation of polydiacetylene.~\cite{BarfordPDA}

The H\"uckel-Hubbard-Ohno model reads
\begin{equation}
\hat{H}_{\rm el}=\hat{T}+\hat{H}_{\rm int} \; .
\label{eq:Helectronic}
\end{equation}
We must determine the electronic ground-state energy
\begin{equation}
E_{\rm el}\left(\left\{x_l\right\},\left\{y_l\right\}\right)
=\langle \Psi_0 | \hat{H}_{\rm el} | \Psi_0\rangle \; .
\end{equation}
$E_{\rm el}$ parametrically depends on the positions of the atoms. 
For our analytical calculations for the H\"uckel model 
we use periodic boundary conditions
for a ring with $L_C=2L$ atoms, see appendix~\ref{sec:Pei}.

For our numerical investigations for the H\"uckel--Hub\-bard(-Ohno) model, 
we employ open boundary conditions for chains with $L_C=2L+2$ atoms.
The DMRG provides highly accurate results 
for large system sizes with up to $L_C=110$ sites, see appendix~\ref{app:B}.

\subsubsection{Bond compression/stretching energy}

In the adiabatic limit, the energy for bond compression or stretching
parametrically depends 
on the distances between neighboring atoms,
\begin{equation}
E_{\rm CC}\left(\left\{x_l\right\},\left\{y_l\right\}\right)
= \sum_{l=1}^{2L-1} V_{\sigma}(d_l) \; .
\end{equation}
We use a linear force function 
\begin{equation}
V_{\sigma}(r) =K_{\sigma,0} (r-r_0)+\frac{K_{\sigma,1}}{2}(r-r_0)^2  \; ,
\label{eq:Forcefunction}
\end{equation}
where $r_0=1.4\, \hbox{\AA}$ is the average carbon atom 
distance in trans-polyacetylene.
For most of our study we use $K_{\sigma,0} =-4.8\, {\rm eV}/\hbox{\AA}$ and 
$K_{\sigma,1}=42\, {\rm eV}/\hbox{\AA}{}^2$, as motivated in 
Ref.~[\onlinecite{PCCPlegeza}].

\subsubsection{Bond bending energy}

The electronic Hamiltonian and the $\sigma$-bond distortion term
do not lead to a zig-zag geometry 
in the ground state. To stabilize the structure shown 
in Fig.~\ref{fig:structure}
we include the repulsion of the $\sigma$-bonds via a clock spring,
\begin{equation}
E_{\rm CC\, b}\left(\left\{x_l\right\},\left\{y_l\right\}\right)
=\frac{C_{\rm b}}{2} \sum_{l=1}^{2L-1}
\left(\cos(\vartheta_l)-\cos(\Theta_0)\right)^2
\label{eq:defECCbend}
\end{equation}
with 
\begin{eqnarray}
\cos(\pi-\vartheta_l) &=& 
\frac{(x_{l+1}-x_l)(x_l-x_{l-1})}{d_ld_{l-1}} \nonumber \\
&& + \frac{(y_{l+1}-y_l)(y_l-y_{l-1})}{d_ld_{l-1}} \; .
\end{eqnarray}
For the straight chain, we set $C_{\rm b}=0$ and
arrange all atoms on a line, i.e., we set $y_l=0$ from the outset.

To second order in $(\vartheta_l-\Theta_0)$, we may equally write
\begin{equation}
E_{\rm CC\, b}\left(\left\{x_l\right\},\left\{y_l\right\}\right)
=\frac{\widetilde{C}_{\rm b}}{2} \sum_{l=1}^{2L-1}
\left(\vartheta_l-\Theta_0\right)^2
\label{eq:ECCbangles}
\end{equation}
with $\widetilde{C}_{\rm b}=C_{\rm b}\sin^2(\Theta_0)$.
The clock-spring constants differ by a factor $\sin^2(\Theta_0)=3/4$
when we work with angles, as in eq.~(\ref{eq:ECCbangles}),
instead of their cosines, as in eq.~(\ref{eq:defECCbend}).
For the zig-zag chains, we set $\tilde{C}_{\rm b}=3.5\, {\rm eV}/{\rm rad}^2$, i.e.,
$C_{\rm b}=4.667\, {\rm eV}$, which is a reasonable
value for polymers.

\subsubsection{Total energy}

The total energy of the structure is the sum of all three contributions.
We abbreviate the coordinates of the atoms 
in the $l$th unit cell ($l=1,\ldots,L$) by
\begin{equation}
\vec{p}_l=\left(\begin{array}{@{}c@{}}
x_{2l}\\
y_{2l}\\
x_{2l-1}\\
y_{2l-1}
\end{array}
\right)\; .
\end{equation}
Then, the total energy of the structure reads
\begin{eqnarray}
E_{\rm struc}\bigl(\{\vec{p}_l\}\bigr)
&=&
E_{\rm el}\left(\left\{x_n\right\},\left\{y_n\right\}\right)
+E_{\rm CC}\left(\left\{x_n\right\},\left\{y_n\right\}\right) \nonumber \\
&& + E_{\rm CC\, b}\left(\left\{x_l\right\},\left\{y_l\right\}\right) \; .
\label{eq:totalenegery}
\end{eqnarray}
It must be minimized with respect to the positions of the carbon atoms
$\left(\left\{x_n\right\},\left\{y_n\right\}\right)$,
\begin{eqnarray}
E_0&=&E_{\rm struc}\big(\{\vec{R}_{l} \}\bigr) 
\; , \label{eq:E0def}\\
\left.\frac{\partial E_{\rm struc}\big(\{\vec{p}_{l} \}\bigr) }{\partial p_{l,j}}
\right|_{\vec{p}_{l}=\vec{R}_{l}} 
&=& 0 \quad \hbox{for $j=1,\ldots,4$} \; .\label{eq:gradzero}
\end{eqnarray}
By construction, we find the minimum $E_0$ of
the energy functional~$E_{\rm struc}$
at the optimal atomic positions~$\vec{R}_l=(X_{2l},Y_{2l},X_{2l-1},Y_{2l-1})^{\rm T}$ 
for $l=1,\ldots,L$.

\subsection{Optical phonons}

The second derivatives of the ground-state energy with respect to the atomic
positions define the dynamical matrix from which the phonon
spectrum can be calculated.~\cite{Solyom} 
In the following we shall focus
on distortions that are identical in each unit cell (`optical phonons').

\begin{figure}[hb]
\includegraphics[height=5.5cm,angle=90]{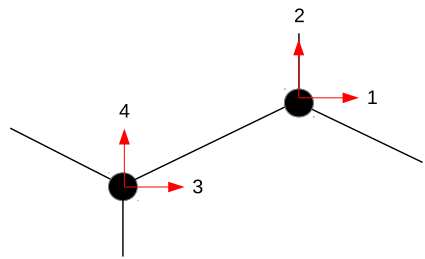}
\caption{Four distortions in the unit cell for optical phonons
in a planar and unflexed zig-zag chain.\label{fig:distortions}}
\end{figure}

\subsubsection{Distortions and dynamical matrix}

By definition, optical phonons are characterized by the fact that the motion
of all atoms is the same when going from one unit cell to the next.
This results from the fact that the light field adds vanishingly small
momentum to the system.~\cite{Solyom}
Thus, we may set
\begin{equation}
\vec{p}_l -\vec{R}_{l}= \vec{\delta} = 
\left( \delta_1,\delta_2,\delta_3,\delta_4\right)^{\rm T} \; .
\end{equation}
The energy in presence of the distortion becomes 
a function of the four parameters $\vec{\delta}$,
\begin{equation}
E(\vec{\delta}) = E_{\rm struc}\big(\{\vec{R}_{l}+\vec{\delta}\}\bigr)\;.
\end{equation}
The distortions are shown schematically in Fig.~\ref{fig:distortions}.

The chain is symmetric under a rotation by 180$^{\circ}$ around its midpoint.
It is useful to work with lattice distortions
that respect this $C_2$-symmetry. Therefore,
for our calculations we henceforth use 
\begin{eqnarray}
\vec{p}_l -\vec{R}_{l}= \vec{\tilde{\delta}} &=& 
\left( 
\tilde{\delta}_1,\tilde{\delta}_2,\tilde{\delta}_3,\tilde{\delta}_4\right)^{\rm T} 
= (\underline{\underline{O}}\cdot \vec{\delta}\,)^{\rm T}
\label{eq:c2symbasis}
\\
&=& 
\left( \frac{\delta_1+\delta_3}{\sqrt{2}},
\frac{\delta_2+\delta_4}{\sqrt{2}},
\frac{\delta_1-\delta_3}{\sqrt{2}},
\frac{\delta_2-\delta_4}{\sqrt{2}}\right)^{\rm T} 
\nonumber 
\end{eqnarray}
with
\begin{equation}
\underline{\underline{O}}^+=\underline{\underline{O}}^{-1}
=\underline{\underline{O}}=\sqrt{\frac{1}{2}} \left(
\begin{array}{rrrr}
1 & 0 & 1 & 0 \\
0 & 1 & 0 & 1 \\
1 & 0 & -1 & 0 \\
0 & 1 & 0 & -1 \\
\end{array}
\right) 
\end{equation}
to separate the symmetry sectors $A_g$ (`gerade')
and $B_u$ (`ungerade').

For the calculation of the dynamical matrix, we need to Taylor expand
the energy to second order,
\begin{equation}
E(\vec{\delta}) \approx  E_0 +\frac{L}{2}
\sum_{i,j=1}^4
K_{i,j} \delta_i \delta_j
\; ,
\end{equation}
where we used eqs.~(\ref{eq:E0def}) and~(\ref{eq:gradzero})
and defined the elements of the real, symmetric dynamical matrix 
\begin{equation}
K_{i,j}=\frac{1}{L}
\sum_{n,m=1}^L 
\left.\frac{\partial^2 E_{\rm struc}\bigl(\{\vec{p}_{l} \}\bigr)}%
{\partial p_{n;i}\partial p_{m;j}}
\right|_{\vec{p}_l=\vec{R}_l}
\label{eq:Kijdef}
\end{equation}
for optical phonons. 
Correspondingly, we have
\begin{equation}
E(\vec{\delta}) \approx  E_0 +\frac{L}{2}
\sum_{i,j=1}^4
\tilde{K}_{i,j} \tilde{\delta}_i \tilde{\delta}_j
\end{equation}
with $\underline{\underline{\tilde{K}}}= \underline{\underline{O}}^+\cdot
\underline{\underline{K}}\cdot \underline{\underline{O}}$.
Note that $\underline{\underline{\tilde{K}}}$ is block diagonal,
\begin{equation}
\underline{\underline{\tilde{K}}}=
\left(
\begin{array}{cc}
\underline{\underline{B_u}} & \underline{\underline{0}} \\
\underline{\underline{0}} & \underline{\underline{A_g}}
\end{array}
\right) \; ,
\end{equation}
where $\underline{\underline{A_g}}$ and $\underline{\underline{B_u}}$,
are $2\times 2$ matrices and $\underline{\underline{0}}$ is the $2\times 2$ 
zero matrix.

\subsubsection{Classical Hamilton function and phonon frequencies}
\label{subsubsec:dynamicmatrix}

The corresponding classical Hamilton function 
for the displacement in one unit cell is given by
\begin{equation}
H_{\rm ph}\bigl(\{\dot{\delta_l}\},\{\delta_l\}\bigr)=
T\bigl(\{\dot{\delta_l}\}\bigr)
+V\bigl(\{\delta_l\}\bigr)
\end{equation}
with 
\begin{eqnarray}
T\bigl(\{\dot{\delta_l}\}\bigr)&=&  \frac{M}{2} \sum_{i=1}^4 (\dot{\delta_i})^2 
= \frac{M}{2} \sum_{i=1}^4 (\dot{\tilde{\delta}}_i)^2 \; , \nonumber \\
V\bigl(\{\delta_l\}\bigr)&=& \frac{1}{2}  \sum_{i,j=1}^4 K_{i,j}\delta_i\delta_j 
=\frac{1}{2}  \sum_{i,j=1}^4 \tilde{K}_{i,j}\tilde{\delta}_i\tilde{\delta}_j 
\; ,
\end{eqnarray}
where $M$ is the mass of the atoms.

The optical phonon frequencies can be derived from
the classical equations of motion.~\cite{Solyom} 
The four phonon frequencies result from the zeros of the characteristic
polynomials $P_{A,B}(\omega^2)$ with
\begin{eqnarray}
P_A(\omega^2)\equiv {\rm Det}\left( -\omega^2 M \underline{\underline{1}}+
\underline{\underline{A_g}}
\right) =0 \;, \nonumber \\
P_B(\omega^2)\equiv {\rm Det}\left( -\omega^2 M \underline{\underline{1}}+
\underline{\underline{B_u}}
\right) =0 \;,
\label{eq:charpol}
\end{eqnarray}
where $M=12u$ ($1u=1.66054\cdot 10^{-27}\, {\rm kg}$) is the mass of an atom
and $\underline{\underline{1}}$ is the $2\times2$ unit matrix.
Since all atoms have equal mass, we immediately find ($l=1,2$)
\begin{equation}
\omega_{l}^{\rm acc}=\sqrt{\frac{\tilde{K}_{B,l}}{M}} \quad , \quad
\omega_{l}^{\rm opt}=\sqrt{\frac{\tilde{K}_{A,l}}{M}} \; , 
\label{eq:frequ}
\end{equation}
where $\tilde{K}_{A/B,l}$ are the two eigenvalues of the dynamical 
matrices~$\underline{\underline{A_g}}$
and $\underline{\underline{B_u}}$, respectively.

For periodic boundary conditions, the two eigenvalues 
in the $B_u$ symmetry sector are zero, $\tilde{K}_{B,1}=\tilde{K}_{B,2}=0$, 
because they correspond
to a horizontal or vertical motion of the whole chain.
For our numerical investigations, we fix the first and last atom so that
$\tilde{K}_{B,1}$ and $\tilde{K}_{B,2}$ are not exactly zero. 
This gives rise to two `acoustic' phonons in our DMRG calculations.
Their energies are proportional to the inverse of the total mass of the chain
so that the energy of the acoustic modes
vanishes in the thermodynamic limit, $\omega_l^{\rm acc}\sim\sqrt{1/L_C}$.
Differences between the analytic and numerical results for
$\omega_{1}^{\rm opt}$ and $\omega_{2}^{\rm opt}$ in the H\"uckel model
can be used to assess the importance of finite-size effects and to test
the numerical accuracy of our approach.

When we measure energies in eV and distances in~\AA, 
the entries of the dynamical matrix have the unit ${\rm eV}/\hbox{\AA}{}^2$.
Thus, the phonon frequencies in eq.~(\ref{eq:frequ})
are given in units of $\sqrt{{\rm eV}/u}/\hbox{\AA}$.
To express the phonon frequencies in terms of wavenumbers (cm$^{-1}$), we use
the conversion factor
\begin{equation}
\frac{\lambda^{-1}}{{\rm cm}^{-1}}
=\frac{\omega}{\sqrt{{\rm eV}/u}/\hbox{\AA}}
\sqrt{\frac{1.60219}{1.66054}}\cdot 10^{14} \frac{1}{2\pi\,  2.997925\cdot 10^{10}}\; .
\label{eq:Florianfactor}
\end{equation}
The conversion factor amounts to 521.473.

\section{H\"uckel model}
\label{sec:Hueckelmodel}

We start the presentation of our results with an analysis of the
H\"uckel model that can be solved analytically for periodic boundary conditions.
Therefore, we can assess the quality of the numerical DMRG calculations, i.e.,
we study edge effects and finite-size effects for the average bond length and
the dimerization, and the system-size dependence of the optical phonon frequencies.
We shall show that DMRG provides 
a reliable description for finite systems,
and we can safely extrapolate the optical phonon data  to the thermodynamic limit
if the single-particle gap is converged for the maximal system sizes
that we have reached numerically.

\subsection{Average bond length and dimerization}

For non-interacting electrons,
the average bond length and dimerization
are the same for the straight and zig-zag chains 
because the energy is solely a function
of the bond lengths so that bond angles are irrelevant.

\begin{figure}[b]
\hspace*{-20pt}
\includegraphics[width=7.9cm]{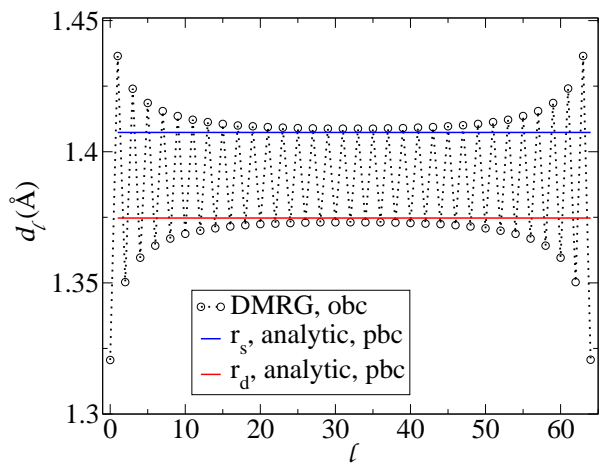}
\caption{Bond lengths for the H\"uckel model on a straight chain 
with $t_0=2.5\, {\rm eV}$, 
$\alpha=4.0\, {\rm eV}/\hbox{\AA}$, 
$K_{\sigma,0} =-4.8\, {\rm eV}/\hbox{\AA}$, and 
$K_{\sigma,1}=42\, {\rm eV}/\hbox{\AA}{}^2$ 
for $L=32$ unit cells for periodic boundary
conditions (pbc, analytic calculation, $L_C=64$) and open boundary 
conditions (obc, DMRG, $L_C=66$).\label{fig:endeffectP}}
\end{figure}

\subsubsection{Edge effects}

In Fig.~\ref{fig:endeffectP} we show the bond length $d_l$
as a function of the bond coordinate~$l$ for the H\"uckel model
on a straight chain with $L=32$ unit cells
for periodic boundary conditions (analytic result, pbc)
and for open boundary conditions (DMRG, obc).
As seen from the figure,
the lengths of the single and double bonds 
obtained from open boundary conditions
agree within a small error margin
with the analytical result for periodic boundary conditions
not only in the center of the chain but for all sites $20<l<46$.
Therefore, our calculated average length~$\overline{r}$
and dimerization~$\Delta$,
taken at the middle of the chain,
are not influenced by edge effects for moderately large chains, $L\gtrsim 30$.

\begin{figure}[hb]
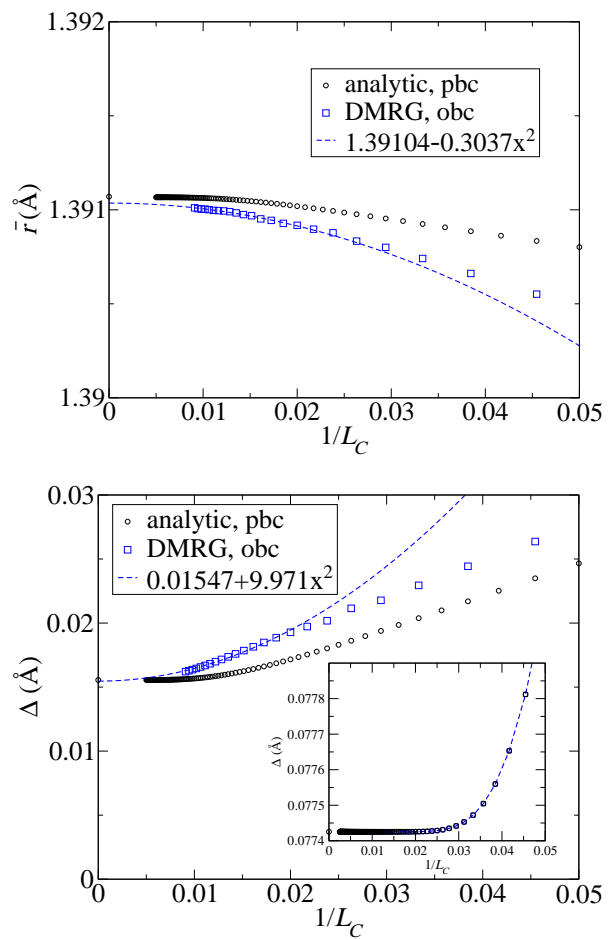

\includegraphics[width=7.9cm]{averager-Hueckel-smallgap.eps}\\[9pt]
\includegraphics[width=7.9cm]{dimer-Hueckel-with-inset.eps}\\
\caption{Average bond length and dimerization for the H\"uckel model 
with $t_0=2.5\, {\rm eV}$, 
$\alpha=4.0\, {\rm eV}/\hbox{\AA}$, $K_{\sigma,0} =-4.8\, {\rm eV}/\hbox{\AA}$, and 
$K_{\sigma,1}=42\, {\rm eV}/\hbox{\AA}{}^2$ 
as a function of the inverse chain length for periodic boundary
conditions (pbc, analytic calculation, $L_C=2L$) and for open boundary 
conditions (obc, DMRG, $L_C=2L+2$).
The limiting values for the H\"uckel 
model are $\overline{r}^{\rm H}=1.39107\, \hbox{\AA}$
and $\Delta^{\rm H}=0.01555\, \hbox{\AA}$.
For the quadratic extrapolation of the DMRG data, system sizes $L_C\geq 50$ 
are used.
Inset: Dimerization for $K_{\sigma,1}=29.5\, {\rm eV}/\hbox{\AA}{}^2$.
\label{fig:extrapolate-randDelta}\label{fig:extrapolate-randDelta-zigzag}}
\end{figure}

\subsubsection{Finite-size effects}

The finite-size extrapolation for the average bond length and the dimerization
can nicely be carried out from our DMRG data for up to $L_C=110$,
see Figs.~\ref{fig:extrapolate-randDelta}.
The average bond length is almost independent of system size
and boundary conditions, as seen from the comparison of periodic
boundary conditions (analytical result) and open boundary conditions (DMRG).

Finite-size and interaction effects are more pronounced for the dimerization.
For the H\"uckel model,
open boundary conditions lead to a larger 
dimerization than periodic boundary conditions.
In both cases, the data for finite system sizes can reliably be extrapolated 
to the thermodynamic limit even for 
$K_{\sigma,1}=42\, {\rm eV}/\hbox{\AA}{}^2$ that leads to a small dimerization, 
$\Delta\approx 0.015\, \hbox{\AA}$.

In the inset of Fig.~\ref{fig:extrapolate-randDelta-zigzag}
we show the dimerization for $K_{\sigma,1}=29.5\, {\rm eV}/\hbox{\AA}{}^2$
that leads to a large dimerization, $\Delta\approx 0.077\, \hbox{\AA}$.
The convergence is significantly faster,
and systems as small as $L_C=32$ provide a reliable estimate
for the value in the thermodynamic limit.

\subsection{Optical phonons}

The frequency of the optical phonons sensitively depends 
on the size of the gap. To elucidate this effect, 
we analyze two different parameter sets.

\subsubsection{Large Peierls gap}

We start with a parameter set that leads to a sizable gap and a large dimerization
with weak finite-size dependencies.
For $K_{\sigma,1}=29.5\, {\rm eV}/\hbox{\AA}{}^2$, the average bond length is
$\bar{r}\approx 1.38\, \hbox{\AA}$,  and the dimerization is
$\Delta\approx 0.08\, \hbox{\AA}$, see the inset of Fig.~\ref{fig:extrapolate-randDelta},
close to the values used by Su, Schrieffer, and Heeger.~\cite{SSH}

\begin{figure}[hb]
\includegraphics[width=7.9cm]{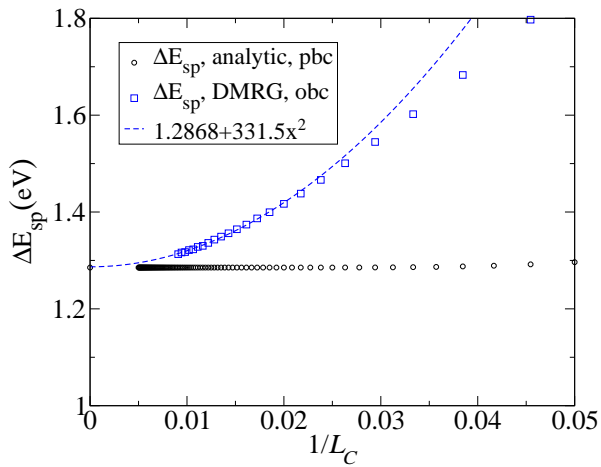}
\caption{Single-particle gap for the H\"uckel model 
with $t_0=2.5\, {\rm eV}$, 
$\alpha=4.0\, {\rm eV}/\hbox{\AA}$, $K_{\sigma,0} =-4.8\, {\rm eV}/\hbox{\AA}$, and 
$K_{\sigma,1}=29.5\, {\rm eV}/\hbox{\AA}{}^2$ 
as a function of the inverse chain length for periodic boundary
conditions (pbc, analytic calculation, $L_C=2L$) and for open boundary 
conditions (obc, DMRG, $L_C=2L+2$).
The limiting value is $\Delta E_{\rm sp}=1.2853\, {\rm eV}$.
For the quadratic extrapolation of the DMRG data, system sizes $L_C\geq 50$ 
are used.\label{fig:gap-largegap}}
\end{figure}

For the H\"uckel model the single-particle gap is given by
\begin{equation}
E_{\rm sp}=2[t(\bar{r}-\Delta)-t(\bar{r}+\Delta)]\; .
\end{equation}
The finite-size dependence of the gap is shown in Fig.~\ref{fig:gap-largegap}.

\begin{figure}[htb]
\includegraphics[width=7.9cm]{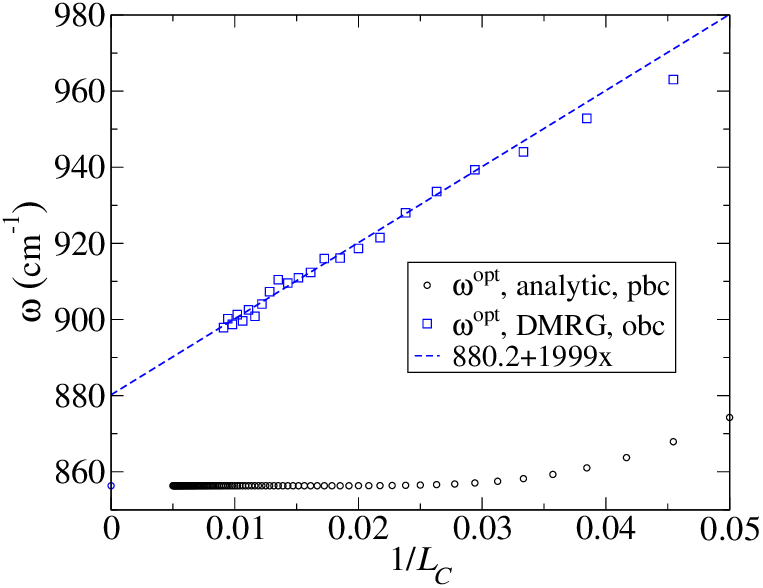}
\caption{Optical phonon frequency for the H\"uckel model on a straight chain
as a function of the inverse chain length
for the same parameter set as in 
Fig.~\protect\ref{fig:gap-largegap}.
The limiting value is $\omega^{\rm opt}=856.3\, {\rm cm}^{-1}$.
For the linear extrapolation of the DMRG data, system sizes $L_C\geq 50$  
are used.\label{fig:phononlargegap}}
\end{figure}

For a large Peierls gap, the finite-size effects are seen to be small
for periodic boundary conditions.
The gap value changes by only 1\% from $L_C=20$ to the thermodynamic limit.
The gap is essentially converged for $L_C\gtrsim 50$.
For open boundary conditions, however,
there are noticeable finite-size effects. Systems as large as $L_C=100$ are 
required to detect the quadratic convergence of the gap as a function of
$1/L_C$.  The extrapolated value agrees with the analytic result
with an accuracy of $0.1\%$.

In Fig.~\ref{fig:phononlargegap} 
we show the phonon frequency for a straight chain
as a function of inverse system size.
The finite-size effects are seen to be small for periodic boundary
conditions. The phonon frequency changes only by 2\% from $L_C=20$ to
the thermodynamic limit. For $L_C\gtrsim 50$, the optical phonon frequency
is essentially converged, as also observed for the single-particle gap.
For open boundary conditions on the contrary, finite-size effects are substantial.
Even for $L_C=110$, the optical phonon frequencies show a fairly
linear dependence on the inverse system size, and a quadratic behavior
on $1/L_C$ is not yet discernible.
As a result, a linear extrapolation of the data to
the thermodynamic limit gives $\omega_{\infty}^{\rm opt}=880\, {\rm cm}^{-1}$,
about $24\, {\rm cm}^{-1}$ or 2.8\% larger than the analytic result
$\omega^{\rm opt}=856.3\, {\rm cm}^{-1}$ for
an infinitely long chain.

We note that the bare optical frequency for a chain without 
electron-phonon coupling ($\alpha=0$) is given by 
$\omega_0=\sqrt{4K_{\sigma,1}/M}=1635\, {\rm cm}^{-1}$.~\cite{supplemental}
For the H\"uckel model, we see a strong renormalization of the optical phonon
frequency. The renormalization of the
phonon frequency is quite large, about a factor of two,
$\omega^{\rm opt}/\omega_0=0.524$.
This behavior reflects the well-known Kohn 
anomaly.~\cite{Kohn,Baeriswyl,supplemental} 

\begin{figure}[t]
\includegraphics[width=7.9cm]{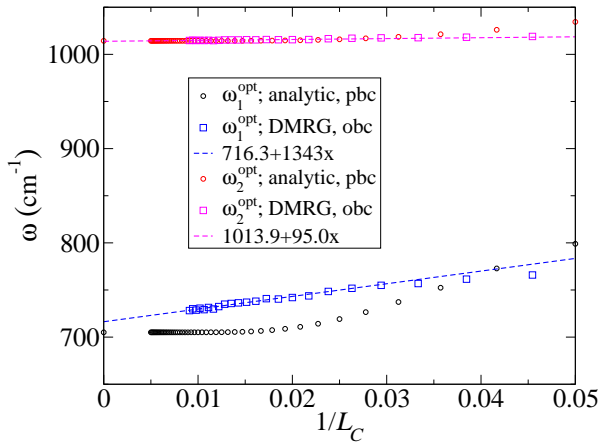}
\caption{Optical phonon frequencies for the H\"uckel model on a zig-zag chain
with $\tilde{C}_b=3.5\, {\rm eV}/{\rm rad}^2$ 
as a function of the inverse chain length
for the same parameter set as in Fig.~\ref{fig:gap-largegap}.
The limiting values are $\omega_1^{\rm opt}=705.1\, {\rm cm}^{-1}$
and $\omega_2^{\rm opt}=1014\, {\rm cm}^{-1}$.
For the linear extrapolation of the DMRG data, system sizes $L_C\geq 50$  
are used.\label{fig:phononszigzaglargegap}}
\end{figure}

\begin{figure}[hbt]
\includegraphics[height=6cm,angle=90]{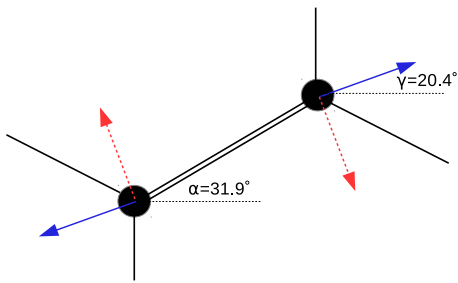}
\caption{Optical phonon distortions associated with
the anti-phase stretching mode at frequency $\omega_1^{\rm opt}$ (blue arrows)
and with 
the anti-phase swinging mode at frequency $\omega_2^{\rm opt}$ (red dotted arrows)
for the H\"uckel model on $L_C=80$ sites and periodic boundary conditions.
for the same parameter set as in 
Fig.~\protect\ref{fig:gap-largegap}.\label{fig:vizLC80Hueckel}}
\end{figure}

In Fig.~\ref{fig:phononszigzaglargegap} 
we show the phonon frequencies for a zig-zag chain
as a function of inverse system size.
The  lower (higher) phonon frequency is associated with
anti-phase stretching (swinging) of the carbon atoms with respective eigenvectors
\begin{eqnarray}
\sqrt{2}\Bigl(\vec{\delta}_1^{\rm opt}\Bigr)^{\rm T}
&=&
\left(\cos(\gamma),\sin(\gamma),-\cos(\gamma), -\sin(\gamma)\right)\; ,
\nonumber\\
\sqrt{2}
\Bigl(\vec{\delta}_2^{\rm opt}\Bigr)^{\rm T}&= &
(\sin(\gamma), -\cos(\gamma),-\sin(\gamma),\cos(\gamma))\; .
\label{eq:vizphononslargegapHueckel}
\end{eqnarray}
For $L_C=80$ sites and periodic boundary conditions we find
$\gamma_{\rm H}=0.355\, {\rm rad}=20.4^{\circ}$. 
They result from the diagonalization of the dynamical matrix with the entries
$\tilde{K}_{33}^{\rm pbc}=24.78$, 
$\tilde{K}_{34}^{\rm pbc}=\tilde{K}_{43}^{\rm pbc}=-7.64$, and
$\tilde{K}_{44}^{\rm pbc}=42.57$ (in units of ${\rm eV}/\hbox{\AA}{}^2$).
The distortions are shown in Fig.~\ref{fig:vizLC80Hueckel}.
The corresponding numbers for open boundary conditions are 
$\tilde{K}_{33}^{\rm obc}=27.48$, 
$\tilde{K}_{34}^{\rm obc}=\tilde{K}_{43}^{\rm obc}=-8.28$,
$\tilde{K}_{44}^{\rm obc}=41.64$ (in units of ${\rm eV}/\hbox{\AA}{}^2$),
and $\gamma_{\rm H}^{\rm obc}=0.431\, {\rm rad}=24.7^{\circ}$.

The anti-phase mode at frequency $\omega_1^{\rm opt}$
corresponds to a stretching of the double bond. Therefore, it is
strongly linked to the $\pi$-electron system and, correspondingly,
it is sensitive to the choice of boundary conditions 
and to the actual size of the single-particle
gap, as already seen for the phonon in the straight chain. Consequently, 
the linear extrapolation 
of the DMRG data to the thermodynamic limit in Fig.~\ref{fig:phononszigzaglargegap}
overestimates the analytic value by about 1.5\%.

The anti-phase swinging mode at frequency $\omega_2^{\rm opt}$
increases the bond-bending energy but barely changes
the kinetic energy of the $\pi$-electrons because the bond lengths remain 
almost constant.
Since $\omega_2^{\rm opt}$ is a very well localized excitation it shows small
finite-size effects for both periodic and open boundary conditions. Correspondingly,
we recover the analytic value in the thermodynamic 
limit from a linear extrapolation of the DMRG data with an accuracy of 0.01\%,
see Fig.~\ref{fig:phononszigzaglargegap}.

In sum, for Peierls insulators with a sizable gap
DMRG calculations for open boundary conditions
can be used to calculate reliably bond lengths, gaps, and
optical phonon frequencies in the thermodynamic limit.

\subsubsection{Small Peierls gap}

\begin{figure}[hb]
\includegraphics[width=7.9cm]{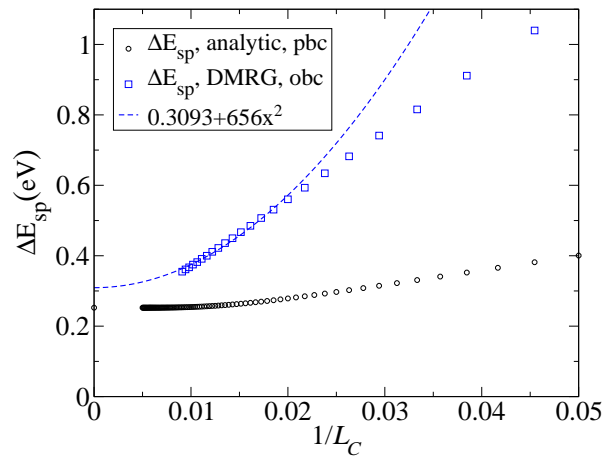}
\caption{Single-particle gap for the H\"uckel model 
with $t_0=2.5\, {\rm eV}$, 
$\alpha=4.0\, {\rm eV}/\hbox{\AA}$, 
$K_{\sigma,0} =-4.8\, {\rm eV}/\hbox{\AA}$, and 
$K_{\sigma,1}=42\, {\rm eV}/\hbox{\AA}{}^2$ 
as a function of the inverse chain length for periodic boundary
conditions (pbc, analytic calculation, $L_C=2L$) and for open boundary 
conditions (obc, DMRG, $L_C=2L+2$).
The limiting value is $\Delta E_{\rm sp}=0.2525\, {\rm eV}$.
For the quadratic extrapolation of the DMRG data, system sizes $L_C\geq 50$  
are used.\label{fig:gap-smallgap}}
\end{figure}

For Peierls insulators with a small single-particle gap,
finite-size effects are much larger and it is
much more difficult to extract optical phonon frequencies 
in the thermodynamic limit from finite-size data.
To illustrate this feature, we analyze a parameter set that leads to a small 
dimerization and a small Peierls gap. 
For $K_{\sigma,1}=42\, {\rm eV}/\hbox{\AA}{}^2$, the average bond length is
$\bar{r}\approx 1.39\, \hbox{\AA}$,  and the dimerization is
$\Delta\approx 0.016\, \hbox{\AA}$, 
see Fig.~\ref{fig:extrapolate-randDelta-zigzag}.
The finite-size dependence of the gap is shown in Fig.~\ref{fig:gap-smallgap}.

For a small Peierls gap, the finite-size effects are seen to be relatively large.
Even for periodic boundary conditions, the gap value for $L_C=20$ is larger than that
in the thermodynamic limit by almost a factor two.
Concomitantly, the finite-size dependence of the gap
is quite large for open boundary conditions. 
A quadratic
dependence of $E_{\rm sp}(L_C)$ on the inverse system size $1/L_C$ 
becomes discernible only for $L_C\gtrsim 100$.
As a consequence, the extrapolated gap is 22\% larger than the analytic value in the
thermodynamic limit.

\begin{figure}[t]
\includegraphics[width=7.9cm]{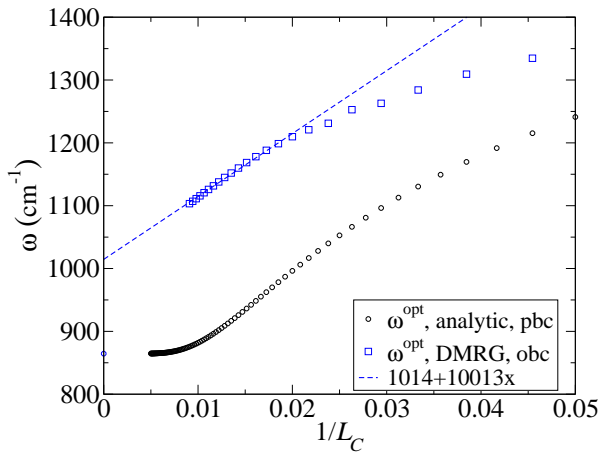}
\caption{Phonon frequency for the H\"uckel model on a straight chain
as a function of inverse chain length
for the same parameter set as in Fig.~\ref{fig:gap-smallgap}.
The limiting value for the H\"uckel model is 
$\omega^{\rm opt}=864.5\, {\rm cm}^{-1}$.
For the linear extrapolation of the DMRG data, system sizes $L_C\geq 50$  
are used.\label{fig:phonlinear}}
\end{figure}

In Fig.~\ref{fig:phonlinear} 
we show the optical phonon frequency for a straight chain
as a function of inverse system size.
The finite-size effects are seen to be large even for periodic boundary conditions.
The phonon frequency at $L_C=20$ is 45\% larger than its value in 
the thermodynamic limit. 
A quadratic dependence of $\omega^{\rm opt}(L_C)$ on the inverse system 
size $1/L_C$ becomes discernible only for fairly large systems,
$L_C\gtrsim 100$. Corresponding to the large finite-size effects,
the boundary conditions also matter. The 
analytic results for periodic boundary conditions and the numerical 
DMRG results
for open boundary conditions differ by 250~cm$^{-1}$
for moderately long chains,~$L_C\approx 100$.
Consequently, the linear extrapolation of the DMRG data to the thermodynamic limit
leads to an optical phonon frequency that
is 17\% or 150 cm$^{-1}$ higher than the analytic value.

\begin{figure}[t]
\includegraphics[width=7.9cm]{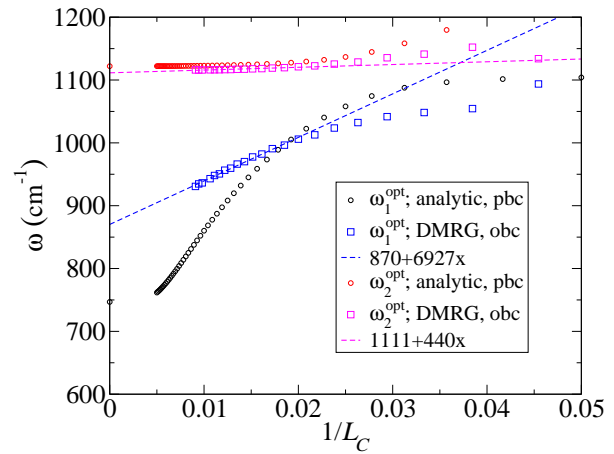}
\caption{Optical phonon frequencies for the H\"uckel model on a zig-zag chain
with $\tilde{C}_b=3.5\, {\rm eV}/{\rm rad}^2$ 
as a function of the inverse chain length
for the same parameter set as in Fig.~\ref{fig:gap-smallgap}.
The limiting values are 
$\omega_1^{\rm opt}=746.9\, {\rm cm}^{-1}$
and
$\omega_2^{\rm opt}=1122\, {\rm cm}^{-1}$.
For the linear extrapolation of the DMRG data, system sizes $L_C\geq 50$  
are used.\label{fig:phonzigzag-hueckel}}
\end{figure}

In Fig.~\ref{fig:phonzigzag-hueckel} we show the optical phonon frequencies 
for a zig-zag chain as a function of inverse system size.
For short chains, the frequency $\omega_1^{\rm opt}$
of the lower-energy optical phonon is almost independent of~$L_C$ until
its frequency becomes comparable to that of the higher-energy phonon 
at $\omega_2^{\rm opt}$ which displays a large finite-size renormalization.
Then, an avoided crossing occurs around $L_C\approx 30$ for our parameter set. 
The frequency $\omega_2^{\rm opt}$ of the higher-energy
phonon levels off and becomes independent of the system size and 
the choice of boundary conditions.
In contrast, the frequency $\omega_1^{\rm opt}$ 
of the lower-energy phonon displays large finite-size effects,
similarly to the optical phonon of the straight chain.
As can be seen from Fig.~\ref{fig:phonzigzag-hueckel}, it drops by
almost 350~cm$^{-1}$ from $L_C=20$ to the thermodynamic limit.
It requires very large systems to determine $\omega_1^{\rm opt}$
from finite-size extrapolations.

The example shows that the Kohn anomaly is clearly visible
in Peierls insulators with a small gap.
The dimerization and single-particle gap display large finite-size effects.
Correspondingly, the optical phonon frequencies show a substantial
finite-size dependence. In the case of the zig-zag chain, an avoided crossing
as a function of system size can be seen.
However, the Kohn anomaly is suppressed in one-dimensional Mott-Peierls insulators
with a sizable single-particle gap, as we shall show in the next section.

\section{H\"uckel-Hubbard-Ohno model}
\label{sec:HHOmodel}

In this section we include the Coulomb interaction. We choose $U=6\, {\rm eV}$
to obtain the average bond length, dimerization, and single-particle gap
as observed for trans-polyacetylene.
Moreover, the long-range Ohno interaction with $V=3\, {\rm eV}$ and static
screening parameter $\epsilon_d=2.3$ 
permit us to reproduce the singlet-exciton binding energy.~\cite{PCCPlegeza}
The Coulomb interaction increases the frequency of the optical phonons 
and, due to the large single-particle gap, it eliminates the signatures of
the Kohn anomaly as seen for small-gap Peierls insulators in the previous section. 

\subsection{Average bond length and dimerization}
\label{sec:HHOdim}

For the H\"uckel-Hubbard(-Ohno) model, 
finite-size and edge effects are effectively suppressed
because the single-particle gap is large in the presence of the Coulomb interaction.
Therefore, the corresponding length scales for the decay of the edge effects
are shorter than for the H\"uckel model, and extrapolations for 
the average bond length and dimerization are even more robust.

For non-interacting electrons and for the H\"uckel-Hub\-bard model
with a purely local interaction, the average bond length and dimerization
are the same for the straight and zig-zag chains because the energy is solely a function
of the bond lengths so that bond angles are irrelevant.
For this reason
we do not discriminate the chain geometry for the H\"uckel-Hubbard model.

\begin{figure}[hb]
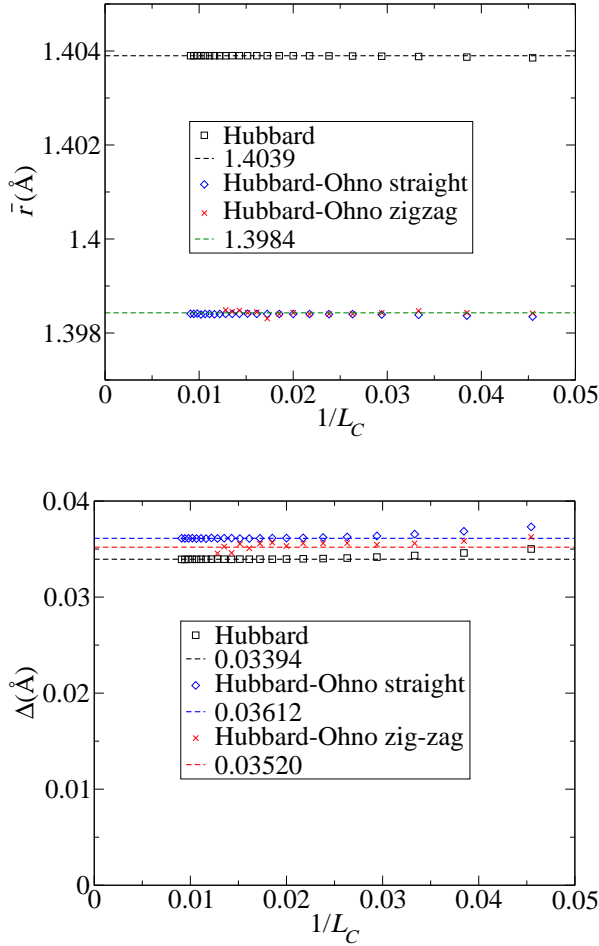

\includegraphics[width=7.9cm]{averager-HHO.eps}\\[18pt]
\includegraphics[width=7.9cm]{dimer-HHO.eps}\\
\caption{Average bond length and dimerization for the 
H\"uckel-Hubbard(-Ohno) model 
with $t_0=2.5\, {\rm eV}$, $U=6\, {\rm eV}$ [$V=3\, {\rm eV}$, 
$\epsilon_d=2.3$],
$\alpha=4.0\, {\rm eV}/\hbox{\AA}$, $K_{\sigma,0} =-4.8\, {\rm eV}/\hbox{\AA}$,
$K_{\sigma,1}=42\, {\rm eV}/\hbox{\AA}{}^2$, and 
$\tilde{C}_b=3.5\, {\rm eV}/{\rm rad}^2$ 
as a function of the inverse chain 
length for open boundary conditions (obc, DMRG, $L_C=2L+2$).
For the constant extrapolation of the DMRG data, system sizes $L_C\geq 50$  
are used.\label{fig:extrapolate-randDeltaHHO}}
\end{figure}

As seen from Fig.~\ref{fig:extrapolate-randDeltaHHO}, 
the influence of the Coulomb interaction on the average bond length 
is very small.
The Hubbard interaction increases the average bond length by
only 0.015~\AA. The long-range Ohno interaction
reduces the average bond length again by a small amount so that the increase
in bond length is below 0.01~\AA\ from the H\"uckel to the
H\"uckel-Hubbard-Ohno model.
Fig.~\ref{fig:extrapolate-randDeltaHHO} also shows that
the zig-zag geometry has a negligible influence
on the average bond length.

In contrast,
the Coulomb interaction is very important for the size of the dimerization.
The dimerization for the H\"uckel-Hubbard(-Ohno) model
with $U=6\, {\rm eV}$ (and $V=3\, {\rm eV}$, $\epsilon_d=2.3$),
is larger than that for the bare H\"uckel model by a factor of two,
from $\Delta^{\rm H}\approx 0.02\, \hbox{\AA}$ 
to $\Delta^{\rm HH}\approx 0.04\, \hbox{\AA}$, the experimental value for
trans-polyacetylene.
As seen from Fig.~\ref{fig:extrapolate-randDeltaHHO},
the main reason for this large increase is the Hubbard interaction
whereas the Ohno contribution is fairly small, below 0.002~\AA. 
For this reason, the chain geometry does not play a big role. As seen from the figure,
the difference between the dimerization for straight and zig-zag chains 
is at most 0.001~\AA. 

Finite-size effects are negligibly small. 
The average bond length is essentially independent
of $L_C$, and the dimerization becomes
independent of system size for $L_C\gtrsim 50$.
Therefore, we fit the DMRG data  to a constant 
in Fig.~\ref{fig:extrapolate-randDeltaHHO}.

\subsection{Optical phonons}

\subsubsection{Linear chain}

\begin{figure}[hb]
\hspace*{-20pt}
\includegraphics[width=7.9cm]{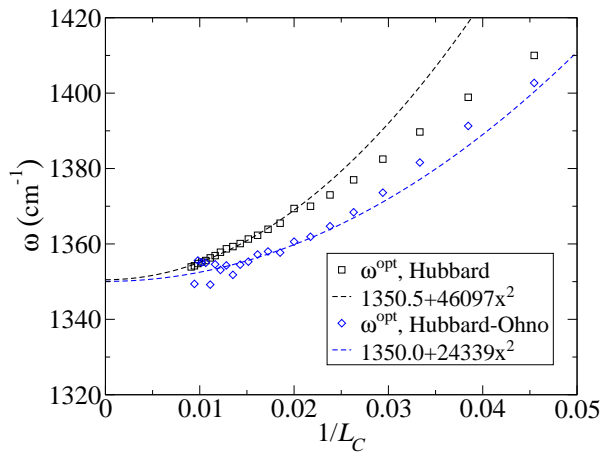}
\caption{Phonon frequency for the H\"uckel-Hubbard(-Ohno) model
on a straight chain as a function of inverse chain length 
for the same parameter set as in Fig.~\ref{fig:extrapolate-randDeltaHHO}.
For the quadratic extrapolation of the DMRG data, system sizes $L_C\geq 50$  
are used.\label{fig:phonlinearHHO}}
\end{figure}

In Fig.~\ref{fig:phonlinearHHO} we show the phonon frequency
as a function of inverse system size for the H\"uckel-Hubbard
and H\"uckel-Hubbard-Ohno models. 
For comparison we note that the bare optical frequency for a chain without 
electron-phonon coupling ($\alpha=0$) is given by 
$\omega_0=\sqrt{4K_{\sigma,1}/M}=1951\, {\rm cm}^{-1}$.~\cite{supplemental}

For our parameter set, the Coulomb interaction leads to a single-particle gap 
of several~eV. As seen from Fig.~\ref{fig:phonlinearHHO},
the phonon frequency shows a very moderate finite-size dependence
and can nicely be extrapolated to the thermodynamic limit.
Moreover, the long-range Ohno interaction shifts the phonon frequency by only
10~cm$^{-1}$, or less, i.e., the frequency shift is almost negligibly small,
below 0.1\%. 
This observation is helpful for parameter optimizations
because DMRG calculations for the H\"uckel-Hubbard model are much less
time consuming than those for the H\"uckel-Hubbard-Ohno model
due to the absence of long range Coulomb interaction.
It was already noted in Ref.~[\onlinecite{PhysRevB.67.245202}]
that the long-range part of the Coulomb interaction barely influences
the effective $\sigma$-bond spring constant.

\subsubsection{Zig-zag chain}
\label{sec:zigzag}

\begin{figure}[htb]
\begin{center}
\includegraphics[width=8.6cm]{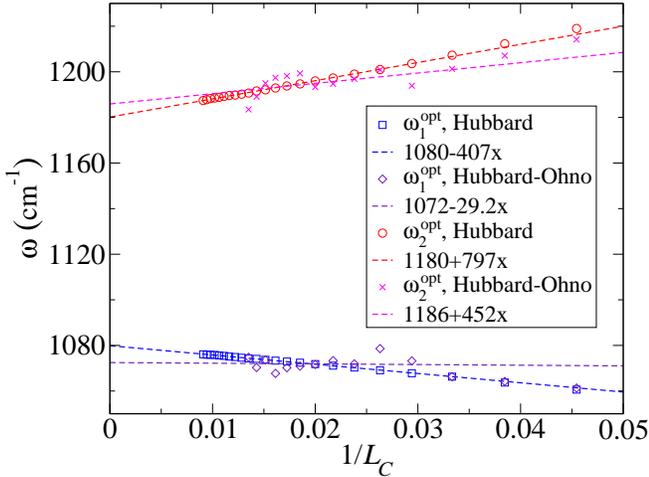}
\caption{Phonon frequencies for the H\"uckel-Hubbard(-Ohno) model
on a zig-zag chain as a function of inverse chain length
for the same parameter set as in Fig.~\ref{fig:extrapolate-randDeltaHHO}.
For the linear extrapolation of the DMRG data, system sizes $L_C\geq 30$  
are used.\label{fig:phonzigzagHHO}}
\end{center}
\end{figure}

Lastly, we present results for the zig-zag chain with an angle 
$\Theta_0=120^{\circ}$ between adjacent double and single bonds.
In Fig.~\ref{fig:phonzigzagHHO} we show the frequencies of the 
optical phonons for the H\"uckel-Hubbard-Ohno model.
As also seen for the straight chain in 
Fig.~\ref{fig:phonlinearHHO}, 
the phonon frequencies in the $A_g$ symmetry sector
are shifted upward by the Coulomb interaction. 
As for the straight chain, the long-range Coulomb interaction does not affect
the phonon frequencies significantly. The changes are again of the
order of 10~cm$^{-1}$ or one percent.

The optical phonons are split in energy
by about 100~cm$^{-1}$ for all chain lengths so that 
an anti-crossing of the phonons as a function of frequency
is not observed, in contrast to the bare H\"uckel model with a small gap.
The system sizes $L_C\approx 100$ are still too small to permit a quadratic fit
of the phonon frequencies as a function of system size.
We expect, however, that the linear frequency extrapolations 
lead to the correct phonon frequencies in the
thermodynamic limit, to within 10~cm$^{-1}$ or one percent.

\begin{figure}[htb]
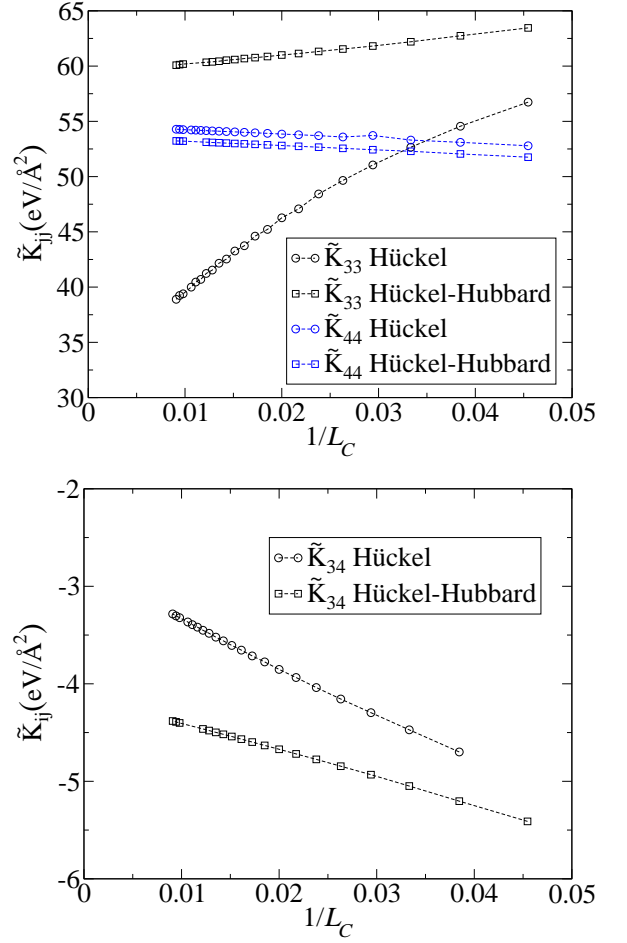

\begin{center}
\mbox{}\\[3pt]
\includegraphics[width=7.8cm]{K33K44.eps}\\[6pt]
\includegraphics[width=7.8cm]{K34.eps}
\caption{Entries of the dynamical matrix $\tilde{K}_{i,j}$ 
in the $A_g$ symmetry sector
for the H\"uckel-Hubbard(-Ohno) model 
with $t_0=2.5\, {\rm eV}$,
$\alpha=4.0\, {\rm eV}/\hbox{\AA}$, $K_{\sigma,0} =-4.8\, {\rm eV}/\hbox{\AA}$,
$K_{\sigma,1}=42\, {\rm eV}/\hbox{\AA}{}^2$,  
$\tilde{C}_b=3.5\, {\rm eV}/{\rm rad}^2$,
[and $U=6\, {\rm eV}$ ($V=3\, {\rm eV}$, $\epsilon_d=2.3$)],
as a function of the inverse chain 
length for open boundary conditions 
(obc, DMRG, $L_C=2L+2$).\label{fig:Kmatrixelements}}
\end{center}
\end{figure}

For completeness,  in Fig.~\ref{fig:Kmatrixelements}
we show the entries of the dynamical matrix $\tilde{K}_{i,j}$
in the $A_g$-sector as a function of system size for the H\"uckel model with a small
gap, and the H\"uckel-Hubbard(-Ohno) model.
In the H\"uckel model with a small gap, 
the matrix element $\tilde{K}_{33}$ for anti-phase
distortions in the $x$-direction displays a large finite-size dependence,
similar to the finite-size gap in Fig.~\ref{fig:gap-smallgap}.
For this reason,  $\tilde{K}_{33}$ equals the value for $\tilde{K}_{44}$ 
around $L_C=30$ which leads to the avoided crossing 
of the phonon frequencies seen in Fig.~\ref{fig:phonzigzag-hueckel}.
This drastic 
finite-size behavior is suppressed by the Hubbard(-Ohno) interaction.
The Coulomb repulsion leads to a much larger single-particle gap that
is well converged as a function of inverse system size for $L_C\gtrsim 50$.
As a consequence, the phonon frequencies shown in  Fig.~\ref{fig:phonzigzagHHO}
are well separated and do not show signatures of an avoided crossing.
Note that $\tilde{K}_{44}$ and $\tilde{K}_{34}=\tilde{K}_{43}$  
are quite similar, i.e., the Coulomb interaction plays a minor role for distortions
that involve the $y$-direction, perpendicular to the chain orientation. 
This is not surprising because in our quasi one-dimensional system 
the electron wave functions are extended only along the $x$-direction. 

\begin{figure}[t]
\begin{center}
\includegraphics[height=6cm,angle=90]{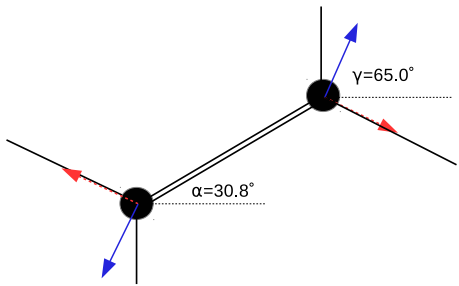}
\end{center}
\caption{Optical phonon distortions associated with
the anti-phase stretching mode at frequency $\omega_2^{\rm opt}$ (red dotted arrows)
and with 
the anti-phase swinging mode at frequency $\omega_1^{\rm opt}$ (blue arrows)
for the H\"uckel-Hubbard model on $L_C=66$ sites
for the same parameter set as in 
Fig.~\protect\ref{fig:extrapolate-randDeltaHHO}.\label{fig:vizHueckelHubbard}}
\end{figure}

In Fig.~\ref{fig:vizHueckelHubbard} we show the eigenvectors for the 
anti-phase oscillations for the H\"uckel-Hubbard model for $U=6\, {\rm eV}$
and $L_C=66$.
The eigenvectors for the
lattice distortions for the H\"uckel-Hubbard(-Ohno) model
are still given by eq.~(\ref{eq:vizphononslargegapHueckel})
but now we have $\gamma_{\rm HH}=1.13\, {\rm rad}=65^{\circ}$.
The eigenvectors result from the diagonalization of the dynamical matrix with the entries
$\tilde{K}_{33}=60.59$, $\tilde{K}_{34}=\tilde{K}_{43}=-4.54$, and
$\tilde{K}_{44}=53.00$ (in units of ${\rm eV}/\hbox{\AA}{}^2$).

A comparison of Fig.~\ref{fig:vizLC80Hueckel} for the H\"uckel model with a large gap
and of Fig.~\ref{fig:vizHueckelHubbard} for the H\"uckel-Hubbard model
shows that the eigenvector pairs are rotated against each other by about 45~degrees.
In the H\"uckel-Hubbard model, there is no clear distinction between 
a `stretching' and a `swinging' mode. 
Instead, the phonon mode with the (higher)
energy $\omega_2^{\rm opt}$ corresponds
to a stretching of the single bond whereas the phonon mode with the (lower)
energy $\omega_1^{\rm opt}$ involves a stretching (and swinging) of the double bond.
Therefore, both modes are very similar in their finite-size behavior.

\section{Summary and conclusions}
\label{sec:summconc}

In this work, we calculated the optical phonon frequencies for 
the H\"uckel and H\"uckel-Hubbard(-Ohno) models on linear and zig-zag chains
with up to $L_C=110$ sites 
using the Density-Matrix Renormalization-Group (DMRG) method.
When the electron-electron interaction is absent (H\"uckel model),
the (optical) phonon spectrum 
can be calculated analytically for periodic boundary
conditions and thus provides a benchmark test for the DMRG calculations.
For systems with a large single-particle gap, 
the analytic and numerical results 
for the average bond length, the dimerization, the single-particle gap,
and the optical phonon frequencies 
agree very well, which validates the applicability of the DMRG approach.

When the Peierls gap is small, the Kohn anomaly leads to
an avoided crossing of the two optical phonon branches 
of the zig-zag chain as a function of inverse system size.
Numerically, it is difficult to reach system sizes where the frequency of
the bond-stretching phonon becomes close to its value in the thermodynamic limit.
Moreover, we treated the lattice deformations classically.
This is justified in presence of a large single-particle gap but
causes problems when the system is close to the Peierls 
transition. In systems with a small Peierls gap, the phonons ought to be treated 
quantum mechanically.~\cite{PhysRevB.83.195105}

In presence of a sizable Hubbard interaction, of the order of half the bandwidth,
the gap for single-particle excitations is large and the Kohn anomaly is suppressed.
Finite-size effects of the average bond length, the dimerization, the single-particle gap, 
and the optical phonon frequencies are small, and the DMRG results
can be extrapolated reliably to the thermodynamic limit.
For the zig-zag chain we find that the two optical phonon
modes are energetically well separated for all system sizes. 
We find that a moderate Ohno repulsion increases the single-particle gap 
whereas it barely influences the average bond length, the 
dimerization, and the optical phonon frequencies.
When we increase the interaction parameters 
we observe that a larger Hubbard-$U$
and/or a larger Ohno-$V$ increase both the band gap and the dimerization 
substantially.
Contrary to this, the average bond length remains insensitive to the Coulomb interaction.

The H\"uckel-Hubbard-Ohno model on a zig-zag chain includes 
the basic structural elements and fundamental electronic components
for the description of optical phonons in trans-polyacetylene.
Raman spectroscopy reveals four $A_g$ phonon 
modes~\cite{Kuzmany,Takeuchi1,Takeuchi2,Ehrenfreund} of which two
display a pronounced dependence on the chain length.~\cite{Kuzmany}
Their frequencies are $\omega_{\rm A}\approx 1070\, {\rm cm}^{-1}$ and
$\omega_{\rm C}\approx 1460\, {\rm cm}^{-1}$. Apparently,
$\omega_A$ is close to $\omega_{1}^{\rm opt}\approx 1080\, {\rm cm}^{-1}$, 
as derived for the H\"uckel-Hubbard-Ohno model.
Even $\omega_C$ is in the range of 
$\omega_{2}^{\rm opt}\approx 1180\, {\rm cm}^{-1}$.
Note, however, that the modes~A and~C seen in experiment couple to the 
motion of the hydrogen atoms whereas we studied solely the motion of
the carbon atoms. Therefore, a realistic description of optical phonons 
requires the inclusion of the hydrogen atoms.
Work in this direction is in progress.

\appendix

\section{Non-interacting electrons}
\label{sec:Pei}

In this appendix, we provide analytical expressions for the ground-state conformation
and optical phonons for non-interacting electrons on straight and zig-zag chains
with periodic boundary conditions. 

\subsection{Ground-state energy and bond lengths}

The operator for the kinetic energy reads
\begin{eqnarray}
\hat{T}&=& -\sum_{\sigma}\sum_{n=1}^L t_d
\left( 
\hat{c}_{2n-1,\sigma}^+\hat{c}_{2n,\sigma}^{\vphantom{+}} 
+
\hat{c}_{2n,\sigma}^+\hat{c}_{2n-1,\sigma}^{\vphantom{+}} 
\right)\nonumber \\
&&-\sum_{\sigma}\sum_{n=1}^L t_s
\left( 
\hat{c}_{2n,\sigma}^+\hat{c}_{2n+1,\sigma}^{\vphantom{+}} 
+
\hat{c}_{2n+1,\sigma}^+\hat{c}_{2n,\sigma}^{\vphantom{+}} 
\right) \, ,
\end{eqnarray}
where 
\begin{eqnarray}
t_d&=&t_0\exp\left(-(r_d-r_0)\alpha/t_0\right) \; , \nonumber \\
t_s&=&t_0\exp\left(-(r_s-r_0)\alpha/t_0\right)
\label{eq:tdts}
\end{eqnarray}
are the electron transfer matrix elements for short bonds (atomic distance $r_d$)
and long bonds (atomic distance $r_s$).
The length of the bonds is modulated periodically due to the Peierls effect.

The Hamiltonian is readily diagonalized in momentum space,~\cite{supplemental} 
\begin{eqnarray}
\hat{T}&=& \sum_{k,\sigma} E(k) \left( 
\hat{a}_{k,\sigma,+}^{\dagger}\hat{a}_{k,\sigma,+}^{\vphantom{\dagger}}
-
\hat{a}_{k,\sigma,-}^{\dagger}\hat{a}_{k,\sigma,-}^{\vphantom{\dagger}}
\right)\; , \nonumber \\
E(k_m)&=& \sqrt{
(t_s+t_d)^2\cos^2(k_m) +(t_s-t_d)^2\sin^2(k_m)} \; , 
\nonumber \\
k_m&=&(\pi/L) m\, , \, m=-(L/2)+1,\ldots,(L/2) \; .
\label{eq:Ekm}
\end{eqnarray}
The kinetic energy per unit cell as a function of $r_s$ and $r_d$ is given by
\begin{equation}
T(r_s,r_d)=-\frac{2}{L} \sum_{m=-L/2+1}^{L/2} E(k_m) \; .
\label{eq:Tperunitcell}
\end{equation}
The parameters $r_d$ and $r_s$ follow from the minimization of the kinetic energy
and the potential energy per unit cell.

The potential energy is given by the compression energy per unit cell,
\begin{eqnarray}
E_{\rm CC}(r_s,r_d)&=&V_{\sigma}(r_s)+V_{\sigma}(r_d) \; , \nonumber \\
V_{\sigma}(r)&=&K_{\sigma,0}(r-r_0)+\frac{K_{\sigma,1}}{2}(r-r_0)^2 \; ,
\end{eqnarray}
see eq.~(\ref{eq:Forcefunction}).
The optimal values $(R_s,R_d)$ for the bond lengths
follow from the (numerical) solution of the equations
\begin{eqnarray}
V_{\sigma}'(R_s)=\left.\frac{\partial T(r_s,r_d)}{\partial r_s}\right|_{r_s=R_s,r_d=R_d}
\nonumber \; , \\
V_{\sigma}'(R_d)=\left.\frac{\partial T(r_s,r_d)}{\partial r_d}\right|_{r_s=R_s,r_d=R_d}
\; .
\label{eq:getmeRsRd}
\end{eqnarray}
Note that $(R_s,R_d)$ depend on the number of unit cells~$L$ of the chain.

We rewrite the total energy in terms of the
average bond length and the dimerization using the dimensionless variables
\begin{eqnarray}
s&=&\frac{\alpha(r_d+r_s-2r_0)}{2t_0}\nonumber \; ,\\
v&=&\frac{\alpha(r_s-r_d)}{2t_0}\; .
\label{eq:give-s-and-v}
\end{eqnarray}
Then, the total energy per unit cell becomes
\begin{eqnarray}
E/L&=& -4 t_0e^{-s}\cosh(v)\nonumber \\
&& \times \frac{1}{L} \sum_{|k|\leq \pi/2}
\sqrt{\cos^2(k)+\tanh^2(v)\sin^2(k)} \nonumber \\
&& + \frac{2K_{\sigma,0} t_0}{\alpha}s +\frac{K_{\sigma, 1}t_0^2}{\alpha^2}
(s^2+v^2) \; .
\end{eqnarray}
At the optimal values  $s_0$ and $v_0$
the gradient of the energy vanishes. This leads to the coupled equations
\begin{eqnarray}
0 &=& \frac{2K_{\sigma,0} t_0}{\alpha} +\frac{2K_{\sigma, 1}t_0^2}{\alpha^2}s_0
\nonumber \\
&& +4t_0e^{-s_0} \cosh(v_0)\nonumber \\
&& \times \frac{1}{L} \sum_{|k|\leq \pi/2}
\sqrt{\cos^2(k)+\tanh^2(v_0)\sin^2(k)} \nonumber \; , \\
0 &=& \frac{2K_{\sigma, 1}t_0^2}{\alpha^2}v_0 \\
&& - 4t_0e^{-s_0} \frac{1}{L} \sum_{|k|\leq \pi/2}
\frac{\sinh(2v_0) }{\sqrt{2\cos(2k)+2\cosh(2v_0)}} \; ,\nonumber 
\end{eqnarray}
which thus define $R_s$ and $R_d$  from eq.~(\ref{eq:give-s-and-v}).
For non-interacting electrons,  the bond lengths are independent of 
the bond angles so that the above expressions apply for both straight
and zig-zag chains. The expressions are evaluated using 
{\sc Mathematica}.~\cite{Mathematica}

\subsection{Optical phonons}

\subsubsection{Straight chain}
\label{subsec:HMstraight}

For an optical phonon, the displacements leave the length of the unit
cell invariant, $r_d+r_s=R_d+R_s=a_0$. The optical
phonon corresponds to an anti-phase oscillation of the atoms in the unit cell.
Its effective spring constant is given by
\begin{equation}
K_{\rm eff}=\frac{\alpha^2}{2t_0^2}
\left.\frac{\partial^2 (E/L)}{\partial v^2}\right|_{s=s_0,v=v_0} \; ,
\end{equation}
where we included the Jacobi determinant of the  
transformation~(\ref{eq:give-s-and-v}). Explicitly,
\begin{eqnarray}
K_{\rm eff}&=&K_{\sigma,1}-\frac{\alpha^2}{t_0}e^{-s_0}\frac{1}{L} \sum_{|k|\leq \pi/2}
H(k)
\nonumber \\
H(k)&=&
\frac{3+4\cos(2k)\cosh(2v_0)+\cosh(4v_0)}{\sqrt{2}(\cos(2k)+\cosh(2v_0))^{3/2}} \; ,
\end{eqnarray}
and 
\begin{equation}
\omega^{\rm opt}=\sqrt{\frac{4K_{\rm eff}}{M}}\; .
\end{equation}
For $t_0=2.5\, {\rm eV}$, $\alpha=4.0\, {\rm eV}/\hbox{\AA}$
$K_{\sigma,0} =-4.8\, {\rm eV}/\hbox{\AA}$,
$K_{\sigma,1}=42\, {\rm eV}/\hbox{\AA}{}^2$, and $M=12u$
we find $\omega^{\rm opt}=864.5\, {\rm cm}^{-1}$
for an infinitely long chain.
For finite chains, the results are shown in Fig.~\ref{fig:phonlinear}.
We compare these results with a bare straight chain without itinerant electrons
for which $\omega^{\rm opt,bare}=\omega_0=\sqrt{4K_{\sigma,1}/M}
=1951.2\, {\rm cm}^{-1}$. Apparently, the renormalization of the
phonon frequency is quite large, about a factor of two,
$\omega^{\rm opt}/\omega_0=0.443$.

\subsubsection{Zig-zag chain}

For the zig-zag chain, the size of the unit cell is given by
\begin{equation}
a_0=\sqrt{R_s^2+R_d^2+R_sR_d}
\end{equation}
because all bonds form an angle of $\Theta_0=2\pi/3$.
Then, the carbon atoms are located at ($n=1,\ldots,L$)
\begin{eqnarray}
X_{2n-1}&=& (n-1)a_0 \;, \nonumber \\
Y_{2n-1}&=&0  \; ,\nonumber \\
X_{2n}&=&(n-1)a_0 +R_d\sqrt{1-\frac{3R_s^2}{4a_0^2}} \nonumber \; ,\\
Y_{2n}&=&\frac{\sqrt{3}R_sR_d}{2a_0}  \; .
\label{eq:groundstateconf}
\end{eqnarray}
Since $(R_s,R_d)$ are determined from eq.~(\ref{eq:getmeRsRd}),
the ground-state conformation is fixed for a given number of unit cells~$L$.

Optical phonons induce displacements of the four atoms in the unit cell
in the directions~$x$ and~$y$. The coordinates in the presence of
the four displacements $\vec{\delta}= (\delta_1,\delta_2,\delta_3,\delta_4)$
are defined by
\begin{eqnarray}
x_{2n-1}&=&X_{2n-1}+\delta_3\; , \nonumber \\
y_{2n-1}&=&Y_{2n-1}+\delta_4\; , \nonumber \\
x_{2n}&=&X_{2n}+\delta_1\; , \nonumber \\
y_{2n}&=&Y_{2n}+\delta_2\; .
\end{eqnarray}
The distortions $\vec{\delta}$ result in a change of the bond lengths.
The distances between neighboring atoms are given by
\begin{eqnarray}
r_s(\vec{\delta})
&=&\sqrt{(x_{2n-1}+a_0-x_{2n})^2+(y_{2n-1}-y_{2n})^2} \; , \nonumber \\
r_d(\vec{\delta})&=&\sqrt{(x_{2n}-x_{2n-1})^2+(y_{2n}-y_{2n-1})^2}
\label{eq:distancesrsandrd}
\end{eqnarray}
for all $1\leq n \leq L$, and we may set $n=1$ for convenience.

The kinetic energy per unit is still given by eq.~(\ref{eq:Tperunitcell}),
\begin{equation}
T\bigl(r_s(\vec{\delta}),r_d(\vec{\delta})\bigr)
=-\frac{2}{L} \sum_{m=-L/2+1}^{L/2} E(k_m) \; ,
\label{eq:Tperunitcellagain}
\end{equation}
where $E(k_m)$ depends on $\vec{\delta}$ via the tunnel amplitudes
$t_{s,d}\bigl(r_s(\vec{\delta}),r_d(\vec{\delta})\bigr),$
see eqs.~(\ref{eq:tdts}),~(\ref{eq:Ekm}).
The bond-length distortions result in the potential-energy contributions
\begin{equation}
E_{\rm CC}(\vec{\delta})= V_{\sigma}(r_s(\vec{\delta}))
+ V_{\sigma}(r_d(\vec{\delta}))
\end{equation}
and 
\begin{equation}
E_{\rm CC\, b}(\vec{\delta})= C_{\rm b} \left[\cos\Bigl(\vartheta(\vec{\delta})\Bigr)
-\cos(\Theta_0)\right]^2
\end{equation}
per unit cell with
\begin{eqnarray}
-\cos\left(\vartheta(\vec{\delta})\right)&=&
\frac{(a_0-X_2+\delta_3-\delta_1)(X_2+\delta_1-\delta_3)
}{r_s(\vec{\delta})r_d(\vec{\delta})}\nonumber \\[3pt]
&&+\frac{(\delta_4-Y_2-\delta_2)(Y_2+\delta_2-\delta_4)
}{r_s(\vec{\delta})r_d(\vec{\delta})} \; .
\label{eq:cosvartheta}
\end{eqnarray}
The total energy per unit cell becomes
\begin{equation}
E_{\rm struc}(\vec{\delta})=
T\bigl(r_s(\vec{\delta}),r_d(\vec{\delta})\bigr)
+
E_{\rm CC}(\vec{\delta})+E_{\rm CC\, b}(\vec{\delta}) \; .
\end{equation}
By construction, see eq.~(\ref{eq:getmeRsRd}), the gradient of
$E_{\rm struc}(\vec{\delta})$
vanishes at $\vec{\delta}=\vec{0}$,
as it must be for a stable ground-state conformation.

The entries of the dynamical matrix $\underline{\underline{K}}$
are determined from
\begin{equation}
K_{i,j}= \left.
\frac{\partial^2E_{\rm struc}(\vec{\delta})}{\partial \delta_i\partial \delta_j} 
\right|_{\vec{\delta}=\vec{0}}
\; .
\end{equation}
The dynamical matrix is given in units of ${\rm eV}/\hbox{\AA}{}^2$.
As seen from eqs.~(\ref{eq:distancesrsandrd}) and~(\ref{eq:cosvartheta}),
only the combinations $\delta_1-\delta_3$ and $\delta_2-\delta_4$ 
appear in the energy. Therefore,
two of the eigenvalues of the dynamical matrix are zero, as required for optical phonons.
The two zero eigenvalues correspond to the motion of the chain as a whole
into the $x$-direction [displacement eigenvector $\vec{\delta}_x=(1,0,1,0)$]
and in the $y$-direction 
[displacement eigenvector $\vec{\delta}_y=(0,1,0,1)$].
We verify numerically that the other two eigenvalues are positive, 
as it must be for a stable ground-state configuration.

\section{Dynamical matrix from DMRG calculations}
\label{app:B}

In this appendix, we provide some details of our DMRG algorithm, we discuss how
the ground-state conformation is obtained iteratively, and we show how
to calculate the elements of the dynamical matrix straightforwardly
using DMRG.

\subsection{DMRG algorithm}

We investigate the H\"uckel, the H\"uckel-Hubbard, and 
the H\"uckel-Hubbard-Ohno models 
with open boundary condition applying the density-matrix renormalization group 
(DMRG) method.~\cite{White-1992,White-1993}
We perform simulations on system sizes 
from $L_{\rm C}=10$ up to $L_{\rm C}=110$ 
($L_{\rm C}=2L+2$) in steps $\Delta L_{\rm C}=4$.
The precision of the calculations is controlled in terms of the 
dynamic block-state selection (DBSS) approach,~\cite{Legeza-2003,Legeza-2004} 
whereby we keep up to 1000~block states and perform six sweeps.

Using the DMRG as a kernel, we implement a self-consistent 
geometrical optimization method in order to obtain the relaxed geometry, 
i.e., the geometry with the lowest ground-state energy.
In each iteration step the DMRG solves the electronic Hamiltonian 
problem $\hat{H}_{\rm el}$~(\ref{eq:Helectronic})
for a fixed atomic conformation. Moreover, the algorithm provides
the transition and occupation probabilities in~(\ref{eq:kinetic-operator}) 
and (\ref{eq:coulomb-operator}).
Using these expectation values for the construction of the electronic energy 
term~(\ref{eq:Helectronic}), the total energy~(\ref{eq:totalenegery}) 
becomes a function of the atom coordinates 
$\left(\left\{x_n\right\},\left\{y_n\right\}\right)$ that is
minimized using a gradient search.

\subsection{Optimization of the ground-state structure}

The minimization of the total energy of the structure, 
eq.~(\ref{eq:totalenegery}), is achieved iteratively.
For a fast convergence, an educated guess for the ground state
conformation is helpful.
We start with the homogeneous conformation investigating small system size 
with $L=2$ unit cells.
We set $r_{{\rm s},l}=r_{{\rm d},l}=r_0=1.4\, \hbox{\AA}$ 
in eq.~(\ref{eq:groundstateconf}) so that
$a_0=\sqrt{3}r_0$ is the bare size of the unit cell. 
In order to speed up convergence, for larger systems with $L>2$ unit cells a distorted 
initial geometry is constructed utilizing the optimized geometry of system with $L-2$ 
unit cells.
We set $r_{{\rm s},l}=r_{{\rm d},l}=r_0=1.4\, \hbox{\AA}$ 
in eq.~(\ref{eq:groundstateconf}) so that
$a_0=\sqrt{3}r_0$ is the bare size of the unit cell. 

\subsubsection{Iteration}

The algorithm seeks for the self-consistent solution of
the structural and the electronic problem:
\begin{enumerate}
\item The structure determines the parameters for the
$\pi$-electrons' nearest-neighbor transfer 
and their mutual Coulomb interaction;
\item The potential energy landscape shaped by the $\sigma$-bonds and
the $\pi$-electrons defines the structure.
\end{enumerate}

Correspondingly, the algorithm proceeds as follows.
\begin{enumerate}
\item We define for the $k$th iteration ($k=1,2,\ldots, k_{\rm max})$
for $n=1,2,\ldots,2L$
\begin{equation}
\vec{r}_n^{\,(k)} = 
\left( \begin{array}{c}
x_n^{(k)}
\\
y_n^{(k)}
\end{array}
\right) \; .
\end{equation}
Typically, $k_{\rm max}=10$ is sufficient to obtain convergence.
\item 
The $k$th DMRG run is based on the atomic positions
$\vec{r}_n^{\,(k-1)}$ $(k=1,2,\ldots,k_{\rm max}$).
It provides the elements of the single-particle density matrix 
for nearest neighbors ($n=1,2,\ldots,2L-1)$
\begin{equation}
P_{n,\sigma}^{(k)}\equiv
\langle 
\hat{c}_{n,\sigma}^{\dagger}\hat{c}_{n+1,\sigma}^{\vphantom{\dagger}} 
+
\hat{c}_{n+1,\sigma}^{\dagger}\hat{c}_{n,\sigma}^{\vphantom{\dagger}} 
\rangle_0^{(k)} \;,
\end{equation}
the local double occupancy,
\begin{equation}
D_{n}^{(k)}\equiv 
\langle (\hat{n}_{n,\uparrow}-1/2)(\hat{n}_{n,\downarrow}-1/2)\rangle_0^{(k)}
\; ,
\end{equation}
and the elements for the density-density correlation function
($n,j=1,\ldots,2L$, $n\neq j$)
\begin{equation}
C_{nj}^{(k)}\equiv \langle (\hat{n}_n-1)(\hat{n}_j-1)\rangle_0^{(k)}
\; .
\end{equation}
The positions $\vec{r}_n^{\,(k-1)}$ and the
matrix elements $P_{n,\sigma}^{(k)}$, $D_{n}^{(k)}$,
and $C_{nj}^{(k)}$ determine the 
total energy of the structure $E_{\rm struc}^{(k)}$, see eq.~(\ref{eq:totalenegery}).
\item The iteration cycle stops
if $k=k_{\rm max}$ is reached or if, for $k\geq 2$, 
\begin{equation}
\frac{|E_{\rm struc}^{(k-1)}-E_{\rm struc}^{(k)}|}{2L+2}<\epsilon
\end{equation}
with $\epsilon=10^{-4}$ for a sufficient accuracy.

If neither of the two conditions is fulfilled, we determine new atomic positions.
Starting from the configuration $\vec{r}_n^{\,(k-1)}$ 
the energy functional
$E_{\rm struc}$
for fixed $P_{n,\sigma}^{(k)}$, $D_{n}^{(k)}$, and $C_{nj}^{(k)}$
is minimized with respect to the atomic positions $\vec{r}_n$.
A conjugate gradient method requires
the derivatives of 
$E_{\rm struc}$
with respect to the atomic positions.
The gradients are calculated analytically using the Hellman-Feynman 
theorem.~\cite{Hellmann1,Hellmann2,Feynman}
The corresponding expressions are collected in the supplemental material.

After convergence we find the new positions $\vec{r}^{\,(k)}$ and 
the energetic minimum defines $E_{\rm struc}^{(k)}$.
\end{enumerate}
The steps~2 and~3 are iterated until the iteration cycle stops.

\subsection{Calculation of the dynamical matrix}

\subsubsection{Diagonal terms}
\label{sec:diagonalterms}

We start with the diagonal terms and consider a small distortion
with amplitude $\Delta_i$ in the $i$th component of $\vec{p}_l$
\begin{equation}
e(\Delta_i,L)\equiv 
\frac{E_{\rm struc}\left( \{ \vec{R}_{l}+\Delta_i\vec{e}_i\}\right)-E_0}{L}
=\frac{1}{2}
K_{i,i}\Delta_i^2\; ,
\end{equation}
up to second order in $\Delta_i$.
Thus, we have 
\begin{equation}
K_{i,i}(L)= \lim_{\Delta_i\to 0} \frac{2 e(\Delta_i)}{\Delta_i^2}\; .
\end{equation}
In our approach, we calculate $e(\Delta_i,L)$ for chains with up to 
$L_C=110$ sites
and for three values $\Delta_i^{(k)}$.
Then, we apply a quadratic fit of the parabola
\begin{equation}
\varepsilon(\Delta,L)= (K_{i,i}(L)/2) \Delta^2
\end{equation}
to the three points $\bigl(\Delta^{(k)},e(\Delta^{(k)},L)\bigr)$
and the origin. The curvature defines $K_{i,i}$.

We set 
\begin{equation}
\Delta^{(1)}=0.005\, \hbox{\AA} \; , \; 
\Delta^{(2)}=0.010\, \hbox{\AA} \; , \; 
\Delta^{(3)}=0.015\, \hbox{\AA} \; .
\end{equation}
The choice of these values is derived from typical values
for the displacements. For a phonon of energy $\hbar\omega$, 
the average square displacement at thermal energy $k_{\rm B}T$ is 
$\langle x^2\rangle=k_{\rm B}T/( M\omega^2)$.
For a typical optical phonon energy
of $\hbar\omega=0.2\, {\rm eV}$ in polyacetylene
($1/\lambda=1600\, {\rm cm}^{-1}$), 
the square average displacement for carbon
($M=12u$, $1u=0.9315\, {\rm GeV}/c^2$)
at room temperature ($k_{\rm B}T=0.025\, {\rm eV}$) 
is $\sqrt{\langle x_{\rm C}^2\rangle}\approx 0.015\, \hbox{\AA}$;
for the calculation we used $\hbar c=1974\, {\rm eV}\hbox{\AA}$.

\begin{figure}[t]
\includegraphics[width=8.6cm]{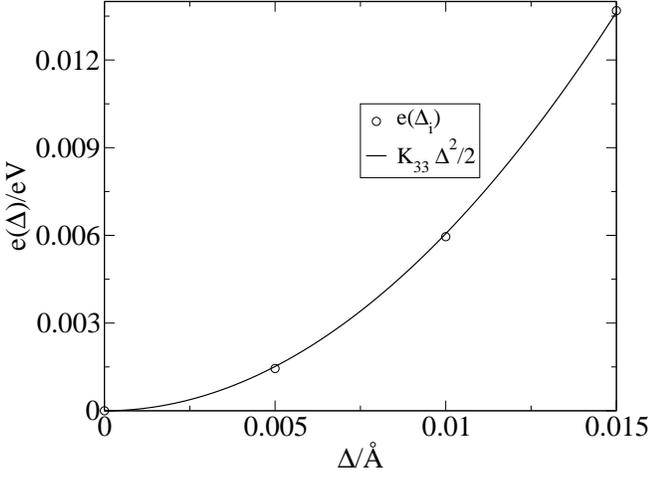}
\caption{Excitation energy for distortions 
of the first carbon atom in each unit cell in the $x$-direction
for the H\"uckel-Hubbard model
on a zig-zag chain with $t_0=2.5\, {\rm eV}$, 
$\alpha=4.0\, {\rm eV}/\hbox{\AA}$, $U=6\, {\rm eV}$,
$K_{\sigma,0} =-4.8\, {\rm eV}/\hbox{\AA}$, and
$K_{\sigma,1}=42\, {\rm eV}/\hbox{\AA}{}^2$ in units of eV
as a function of the distortion $\Delta$ in units of \AA\
for open boundary conditions. The straight line is a parabolic fit
with $K_{33}=60.59\, {\rm eV}/\hbox{\AA}{}^2$.\label{fig:K33fit}}
\end{figure}

As an example,  in Fig.~\ref{fig:K33fit}
we show the extrapolation of $\tilde{K}_{33}$ for the H\"uckel-Hubbard
model for $U=6\, {\rm eV}$ and $L=32$ unit cells on
a zig-zag chain.
It is seen that the extrapolation is stable and provides 
a reliable value for the `spring constant' 
$\tilde{K}_{33}=60.69\, {\rm eV}/\hbox{\AA}{}^2$.
Likewise, we obtain 
$\tilde{K}_{44}=53.00\, {\rm eV}/\hbox{\AA}{}^2$.

\subsubsection{Off-diagonal terms}

For the off-diagonal terms, $1\leq i<j\leq 4$, we consider 
\begin{eqnarray}
e_2(\Delta_i,\Delta_j,L)& \equiv &
E_{\rm struc}\left( 
\{ \vec{R}_{l}+\Delta_i\vec{e}_i+\Delta_j\vec{e}_j\} \right) /L
\nonumber \\
&& -
E_{\rm struc}\left( \{ \vec{R}_{l}+\Delta_i\vec{e}_i\} \right)/L
 \nonumber \\
&& -
E_{\rm struc}\left( \{ \vec{R}_{l}+\Delta_j\vec{e}_j\} \right)/L
+E_0/L \nonumber \\
&=& K_{i,j} \Delta_i\Delta_j 
+{\cal O}\left(\Delta^3\right) \; . 
\label{eq:doubledistort}
\end{eqnarray}

\begin{figure}[ht]
\begin{center}
\mbox{}\\[18pt]
\includegraphics[width=8.6cm]{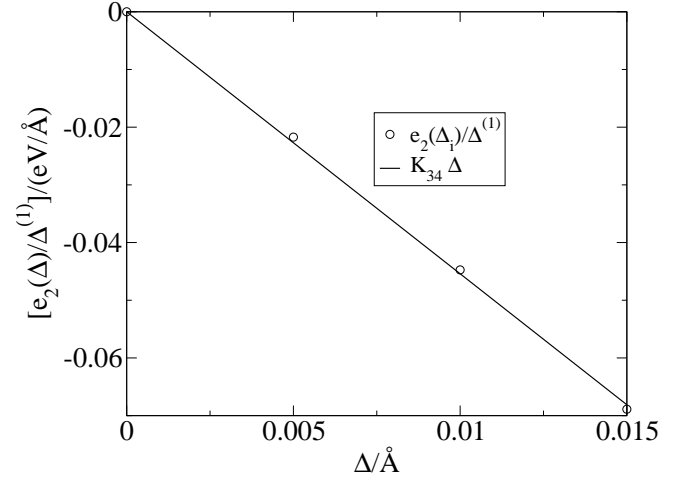}
\caption{Excitation energy for distortions 
of the first carbon atom in each unit cell in the $x$-direction
in the presence of a distortion $\Delta^{(1)}$ in the $y$-direction,
relative to the excitation energies of the individual distortions,
eq.~(\ref{eq:doubledistort}), divided by $\Delta^{(1)}$ in units of eV/\AA.
Data are shown for the H\"uckel-Hubbard model
on a zig-zag chain 
as a function of the distortion $\Delta$ in units of \AA\
for open boundary conditions and
the same parameter set as in Fig.~\ref{fig:K33fit}.
The straight line is a linear fit
with $K_{34}=-4.54\, {\rm eV}/\hbox{\AA}{}^2$.\label{fig:K34fit}}
\end{center}
\end{figure}

We fix $\Delta_i$ at its smallest 
value, $\Delta_i\equiv\Delta^{(1)}$, and find
\begin{equation}
K_{i,j}(L)= \lim_{\Delta_j\to 0} 
\frac{e_2(\Delta^{(1)},\Delta_j,L)}{\Delta^{(1)}}\; .
\end{equation}
In our approach, we calculate $e(\Delta_i^{(1)},\Delta_j^{(k)},L)$ 
for chains with up to $L_C=110$ 
atoms and for three values $\Delta_j^{(1,2,3)}$.
Then, we apply a fit of the linear function
\begin{equation}
\epsilon(\Delta)=K_{i,j}(L)\Delta
\end{equation}
to the three points $\bigl(\Delta_j^{(k)}, 
e(\Delta^{(1)},\Delta_j,L)/\Delta^{(1)}\bigr)$ and the origin.
The slope defines $K_{i,j}(L)$.

As an example, in Fig.~\ref{fig:K34fit}
we show the extrapolation of $\tilde{K}_{34}$ for the H\"uckel-Hubbard
model for $U=6\, {\rm eV}$ and $L=32$ unit cells on
a zig-zag chain.
It is seen again that the extrapolation is fairly stable and provides 
a reliable value for the `spring constant' $\tilde{K}_{34}=\tilde{K}_{43}=
-4.54\,  {\rm eV}/\hbox{\AA}{}^2$.

\subsubsection{Number of DMRG runs}

In this approach, we need a total number 
$N_{\rm run}$ of ground-state DMRG calculations for fixed geometry.
To calculate all elements of the symmetric
$4\times 4$ dynamical matrix, we need
$N_{\rm run}=4\cdot 3 + [(4\cdot 3)/2]\cdot 2 = 12+12=24$ DMRG calculations
for each system size. 

A more compact way to calculate $\tilde{K}_{ij}$ without
an extrapolation in $\Delta$ would be to employ the
second-order Hellman-Feynman theorem. Further information can be found
in the supplemental material.~\cite{supplemental}


\providecommand*{\mcitethebibliography}{\thebibliography}
\csname @ifundefined\endcsname{endmcitethebibliography}
{\let\endmcitethebibliography\endthebibliography}{}

\end{document}